\documentclass[draftclsnofoot,12pt,onecolumn]{IEEEtran}    
\usepackage{color} 
\usepackage{amsmath, amsthm, amssymb}
\usepackage{epsfig}
\usepackage{latexsym}
\usepackage{graphicx} 
\usepackage{cite}
\usepackage{epsfig}
\usepackage{subfigure}
\usepackage{bm}

\title{ Error  Rate and Power Allocation Analysis of Regenerative Networks under Generalized  Fading Conditions}

\author{Mulugeta K. Fikadu, Paschalis C. Sofotasios, Mikko Valkama, Qimei Cui, \\ Sami Muhaidat, and George K. Karagiannidis 

\thanks{M. K. Fikadu and M. Valkama are with the Department of Electronics and Communications Engineering, Tampere University of Technology, 33101, Tampere, Finland \, (e-mail: $\rm \left\lbrace mulugeta.fikadu; mikko.e.valkama \right\rbrace@\rm tut.fi$) }

\thanks{P. C. Sofotasios is with the Department of Electronics and Communications Engineering, Tampere University of Technology, 33101 Tampere, Finland and with the Department of Electrical and Computer Engineering, Aristotle University of Thessaloniki, 54124 Thessaloniki, Greece  \, (e-mail: p.sofotasios@ieee.org)  }

\thanks{ Q.  Cui is with Wireless Technology  Innovation Institute Beijing University of Posts  and Telecommunications  100876, Beijing, China \, (e-mail: cuiqimei@bupt.edu.cn)} 

\thanks{S. Muhaidat is with the Department of Electrical and Computer Engineering, Khalifa University, PO Box 127788, Abu Dhabi, UAE and with the Department of Electronic Engineering, University of Surrey, GU2 7XH, Guildford, U.K. \, (e-mail: muhaidat@ieee.org)}

\thanks{G. K. Karagiannidis is with the Department of Electrical and Computer Engineering, Aristotle University of Thessaloniki, 54124 Thessaloniki, Greece  and with  the Department of Electrical and Computer Engineering, Khalifa University, PO Box 127788, Abu Dhabi, UAE\, (e-mail: geokarag@ieee.org)}
 }
\begin{document}

\maketitle

\begin{abstract}
Cooperative communication has been shown to provide significant increase of transmission reliability and network capacity while expanding coverage in cellular networks. 
 The present work is devoted to the investigation of the end-to-end performance and   power allocation  of a maximum-ratio-combining based regenerative   multi-relay cooperative  network over  non-homogeneous scattering environment, which is  the case in realistic wireless communication scenarios.  
Novel analytic  expressions are derived  for the end-to-end  symbol-error-rate of both $M-$ary Phase-Shift Keying and $M-$ary Quadrature Amplitude Modulation  over   independent and non-identically distributed generalized  fading  channels. 
The offered results are expressed in closed-form  involving the Lauricella function and can be readily evaluated with the aid of  a proposed computational algorithm. Simple expressions are also derived for the corresponding symbol-error-rate at asymptotically high signal-to-noise ratios. 
The derived expressions are corroborated with respective results from computer simulations and are subsequently employed in formulating a power optimization problem that enhances the system performance under  total power constraints within the multi-relay cooperative system.
Furthermore, it is   shown that    optimum power allocation  provides substantial performance gains over  equal power allocation, particularly, when the source-relay and relay-destination paths are highly unbalanced. 
\end{abstract}
 

\section{Introduction}
\indent
Cooperative transmission methods have attracted significant interest over the past decade due to their applicability in size, power, hardware and price constrained devices such as  cellular mobile devices, wireless sensors, ad-hoc networks and  military communications   \cite{R10,R11,R12,  Add_1, Add_4, Add_4c, Add_1b, Add_2b, Add_3b}. 
 Such systems  exploit the broadcast nature and the inherent spatial diversity of wireless paths  and are typically distinguished between  regenerative (decode-and-forward) or non-regenerative (amplify-and-forward) relaying schemes. In general, the digital processing nature of regenerative relaying is considered more efficient than non-regenerative relaying, since the latter typically  requires costly RF transceivers  in order to avoid   forwarding a noisy version of the signal \cite{Add_2, Add_3, Ding_1, Dacosta_1, Maged_1, Maged_2, Himal_1}. \\
\indent
The performance of cooperative systems can be substantially  improved by optimum allocation of the limited overall power to the source and relays of  the network in order to  minimize the overall energy consumption for given end-to-end performance specifications. 
Among others, this can be efficiently achieved by accurately  accounting  for the detrimental effects of multipath fading \cite{ Additional_1, Additional_2, Additional_6, Additional_7, Additional_5, Boss_1, Additional_9, Yacoub_3a,  C:Sofotasios_5, Additional_8, Boss_2, Additional_4, Additional_3, Additional_10, Boss_3, Yacoub_4a, Boss_5, Boss_7, Boss_8, Boss_9, Neo_1, Neo_2} and the references therein.  
Based on this,  the authors in  \cite{R18} derived upper and lower  bounds for the outage probability (OP) of  multi-relay   decode-and-forward (DF) networks over independent but non-identically distributed (i.n.i.d)  Nakagami${-}m$ fading channels.  
In the same context,  the authors in \cite{TD} analyzed the symbol error rate (SER) and OP  of DF  systems with relay selection over i.n.i.d Nakagami${-}m$ fading channels, with integer values of $m$, whereas a  comprehensive analytical framework for  a dual-hop multi-antenna DF   system   under multipath fading  was derived in \cite{SN}.
Likewise, the performance of DF systems over different fading environments was investigated in  \cite{SND, SSI,SSM, Trung_2} whereas analysis for  the SER of dual-hop DF relaying for  $M$-ary phase-shift keying ($M{-}$PSK)  and $M$-ary quadrature amplitude modulation ($M{-}$QAM) over Nakagami$-m$ fading channels was reported in \cite{YM}.   
In addition, optimum power allocation (OPA) in dual-hop regenerative relaying  with respect to pre-defined thresholds for the SER and OP was analyzed in \cite{YM} and \cite{YR}, respectively.
This problem was also addressed in  \cite{R19} and \cite{R20}  for multi-node DF relaying  based on asymptotic SER and for a given   network topology over Rayleigh fading channels, respectively.   
Finally, the authors in \cite{R4}  proposed  power allocation schemes for the case of multi-relay DF communications in the high-SNR regime over  Nakagami${-}m$ fading channels. \\
\indent
Nevertheless, all reported investigations have been carried out in the context of either asymptotic or dual-hop scenarios  as well as considering  only Rayleigh or Nakagami$-m$ fading channels, which is mainly due to the presence of cumbersome integrals that involve combinations of elementary and special functions \cite{C:Sofotasios_2, C:Sofotasios_4, C:Sofotasios_6, C:Sofotasios_7, C:Sofotasios_8, C:Sofotasios_9, Paschalis} and the references therein.   
However, it is recalled that these fading   models are based on the underlying concept of homogeneous scattering environments,  which is not practically realistic  since   surfaces  in most radio propagation environments are spatially correlated \cite{R23}. 
 This issue was addressed in \cite{R1} by proposing the $\eta{-}\mu $ distribution,  which is a generalized fading model that has been shown to provide particularly  accurate fitting to realistic measurement results, while it includes as special cases the well known Rayleigh, Nakagami$-m$ and Hoyt distributions   \cite{R1, R24, Daniel_3, KP}.   
  Based on this, several contributions have been devoted to the analysis of various communication scenarios over generalized fading channels that follow the $\eta-\mu$ distribution, see, e.g., \cite{KP,HY, VA, DM, R7,  WG, R5} and the references therein.   
Motivated by this,  the present work is devoted to the evaluation of the end-to-end  SER in regenerative cooperative communication systems with multiple relays for  $M{-}$PSK and $M{-}$QAM constellations over generalized fading channels    as well as   the corresponding OPA analysis.   Specifically, the contributions of this work are summarized below: 
\begin{itemize}
\item 
Exact closed-form expressions are derived for the end-to-end   SER of $M{-}$PSK and $M{-}$QAM  based multi-relay regenerative networks over generalized multipath fading environments  for  both i.n.i.d and i.i.d scenarios.  
\item
Simple asymptotic  expressions are derived for the above scenarios for   high SNR values. 
\item
The corresponding amount of fading is derived for quantifying the respective fading severity.
\item
Optimal power allocation   based on the convexity of the  derived asymptotic expressions is formulated, to minimize the corresponding SER under sum-power constraint in all nodes.   
\item
The derived expressions are employed in evaluating the performance of the considered system and extracting useful  insights. 
\item
It is shown that a 3dB gain is achieved by OPA for even a small number of relays. 
\item 
It is shown that post-Rayleigh fading conditions result to an  improved performance by  up to 4dB  compared to communications over severe fading channels. 
\item
It is shown that  a  maximum gain of about  $21$dB  occurs, compared to ordinary direct communication, even if only few nodes are employed in non-severe fading conditions. This  renders the resource constrained communication system a meaningful alternative for increasing the quality of service of   demanding emerging wireless  systems. 
\item
A simple   algorithm is proposed for the computation of the generalized Lauricella function. 
\end{itemize} 

The reminder of the present paper is organized as follows:  Section II revisits the considered system and channel models. The exact and asymptotic SER expressions for $M-$QAM and $M-$PSK constellations over generalized multipath fading channels  are derived in Section III.  The optimum power allocation scheme based on  sum-power constraint is provided  in Section IV while Section V  presents the corresponding numerical results along with related discussions and insights. Finally, closing remarks are  given  in Section VI.

\IEEEpubidadjcol


\begin{figure}[tp!]
\centering{\includegraphics[keepaspectratio,width= 3in]{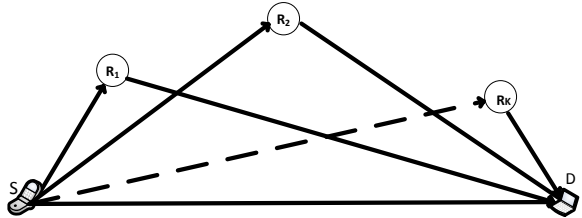}}
\caption{Multi-node dual hop cooperative relay system.} 
\end{figure}

\section{System and Channel Models}

\subsection{System Model }

\indent
We consider a multi-node   cooperative radio access system consisting of a source node $S$,  intermediate relay nodes \textit{$ R_{k} $}, with $k = \{1, 2, \cdots, K\}$, and a destination node $D$, as depicted in Fig.1. Each node in the system is equipped with a single antenna while   a half-duplex decode-and-forward protocol is adopted. Furthermore, in order to avoid inter-relay interference the nodes in the system transmit signals through orthogonal channels.  In phase I, the source broadcasts the signal to the destination and to all relay nodes in the network which yields
 \begin{equation}\label{L1}
 y_{S,D} = \sqrt{P_{0}} \alpha_{S,D} x + n_{S,D}
  \end{equation}
  and
  \begin{equation}\label{L2}
  y_{S,R_k} = \sqrt{P_{0}} \alpha_{S,R_k} x + n_{S,R_k} 
  \end{equation}
where $P_{0}$ is the source transmit power, $x$ is the transmitted symbol with normalized unit energy in the first transmission phase, $ \alpha_{S,D}$ and  $ \alpha_{S,R_k}$ are the complex fading coefficients from the source to the  destination and from the source to the $ k^{\rm th}$ relay, respectively, whereas $ n_{S,D}$ and $n_{S,R_k}$ represent  the corresponding additive-white-Gaussian noise (AWGN) with zero mean and variance $ N_{0} $. 
In the next time slot, if the $k^{\rm th}$ relay decodes correctly, then it forwards  the decoded and re-encoded signal  to the destination with power $  \bar{P}_{R_k} = P_{R_k} $; otherwise, $ \bar{P}_{R_k} = 0 $ yielding
\begin{equation}\label{L3}
y_{R_k,D} = \sqrt{\bar{P}_{R_k}} \alpha_{R_k,D} x + n_{R_k,D} 
\end{equation}
where $\alpha_{R_k,D}$ denotes the complex fading coefficient from the $k^{\rm th}$ relay to the destination and $n_{R_k,D}$ is the corresponding AWGN.
It is also assumed that each   path experiences narrowband multipath fading that follows  the  $\eta-\mu$ distribution and that  MRC diversity scheme is employed at the destination. 
As a result, the combined  output received signal is expressed as follows
\begin{equation}
 y_{D} = w_{0}y_{S,D} + \sum_{k =1}^K w_{k}y_{R_k,D} 
 \end{equation}
where $w_{0} = \sqrt{P_{0}}\alpha^{*}_{S,D}/ N_{0}$  and  $ w_{k} = \sqrt{\bar{P}_{R_k}}\alpha^{*}_{R_k,D}/N_{0}$ denote  the optimal MRC coefficients for   $y_{S,D}$ and $y_{R_k,D}$, respectively with  $(\cdot)^{*}$ representing the complex conjugate operator. 

\subsection{Generalized Multipath Fading Channels}
\indent
It is recalled that    $\eta{-}\mu$ distribution has been shown to account accurately for small-scale variations of the  signal in non-line-of-sight communication scenarios. 
This fading model is  described by the two named  parameters,  $\eta$ and $\mu$, and it is valid for  two different formats that  correspond to two  physical models \cite{R1}.
 The probability density function (PDF) of the instantaneous SNR $ \gamma = \; \mid \alpha  \mid ^{2}P/N_{0} $ is given by  \cite{R1}, \cite{R7}  
\begin{equation}\label{L5}
f_\gamma (\gamma) = \frac {2\sqrt{\pi}\: \mu ^{\mu + \frac{1}{2}} \, h^{\mu} \gamma ^{\mu - \frac{1}{2}}}{\Gamma(\mu)H^{\mu - \frac{1}{2}} \: \overline{\gamma}^{\mu + \frac{1}{2}}}\exp \left( - \frac{2\mu\gamma h}{\overline{\gamma}}\right)I_{\mu -\frac{1}{2}}\left( \frac{2\mu H\gamma}{\overline{\gamma}}\right)
\end{equation}
where  $\overline{\gamma} = E(\gamma) =  (P/N_{0}) \Omega $  is the average SNR per symbol, with $E(\cdot)$ denoting statistical expectation, and $\Omega = E(\mid \alpha \mid ^{2})$ denotes the channel variance. Furthermore, $\Gamma(\cdot)$ and $\: I_{n}(\cdot) $ denote the Euler gamma function and the modified Bessel function of the first kind, respectively \cite{GR}. 
The parameters $h$ and $H$  are given in terms of   $\eta$ in two   formats, as depicted in Table I along with the fading distributions that are included in $\eta-\mu$ as special cases. In terms of physical interpretation  $\eta$ denotes  the scattered-waves power ratio between the in-phase and quadrature components of the scattered waves in each multipath cluster in Format-$1$ and the correlation coefficient  between the in-phase and quadrature components of each multipath cluster in Format-$2$. Likewise, $ \mu =  E^{2}(\gamma)\left( 1 + (H/h\right)^{2})  /2V(\gamma) $ is related to  multipath clustering in both formats, with  $V(\cdot)$ denoting   variance operation, respectively \cite{R1}.

 \begin{table}[tbp!]
\caption{Relation Between $\eta{-}\mu$ Distribution and Other Common Fading Distributions.} 
\centering
\begin{tiny}
\begin{tabular}{|c|c|c|}
\hline
\hline
Fading Distribution & Format-1 & Format-2 
\\
\hline
 $\eta-\mu$  & $h = (1 + \eta^{-1} + \eta)/4,  \, H = (\eta^{-1} - \eta)/4 $ &  $h = 1/(1 - \eta^{2}),  \, H = \eta / (1 - \eta^{2})$  \\
\hline    
Nakagami${-}m$  & $\mu = m , \eta \rightarrow 0 $ or $\eta \rightarrow \infty $   & $\eta\rightarrow \pm 1 $  \\
\cline{2-3}
 & $\mu = m/2, \eta\rightarrow 1 $   & $\eta\rightarrow 0 $   \\
\hline 
 Nakagami${-}q $ (Hoyt)& $\mu = 0.5, \eta = q^{2}$ & $q^{2} = (1-\eta)/ (1 +\eta )$ \\
\hline
 Rayleigh  & $\mu = 0.5, \eta = 1 $ &  $\mu = 0.5 , \eta = 0$  \\
\hline
\hline
\end{tabular}
 \end{tiny} 
 \end{table}

\section{Exact end-to-end Symbol Error Rate Analysis}

The end-to-end SER   for the considered cooperative system can be expressed as    \cite{R19}, \cite{R4} 
\begin{equation}\label{L8}
P_{\rm SER}^{D}  = \sum\limits_{z = 0}^{2^{K}-1} P(e|\mathbf{A} = \mathbf{C}_{z})P(\mathbf{A}= \mathbf{C}_{z})
\end{equation}
where the binary vector   $\mathbf{A}= [A(1),A(2), A(3),. . .,A(K)]$ of dimension $(1\times K)$ denotes the state of the relay nodes in the system, with $A(k)$ taking the binary values of 1 and 0 for successful and unsuccessful decoding, respectively. 
 For the case of statistically independent  channels the joint probability of the possible state outcomes can be represented as  follows:
\begin{equation}
P(\mathbf{A}) = P(A(1))P(A(2))P(A(3)) \cdots P(A(K)) =  \prod_{k = 1}^K P(A(k)). 
\end{equation} 
Furthermore,  $\mathbf{C}_z = [C(1),C(2),C(3),\cdots,C(K)]$ denotes different possible decoding combinations of the relays with $z \in \{0,   2^{K} - 1\}$, where $C(k)$ takes the value of either 0 or 1. The conditional error probability $P(e |\mathbf{A} = \mathbf{C}_{z})$ is the error probability conditioned on   particular decoding results at relays  while $P (\mathbf{A} = \mathbf{C}{_z} )$ is the corresponding joint probability of the   decoding outcomes.  
 Based on the MRC method,  the instantaneous SNR at the destination for given  decoding combination, $\mathbf{C}_{z}$, can be expressed as  \cite{R4} 
 \begin{equation} \label{MGF} 
 \gamma_{\text{MRC}} (\mathbf{C}_{z})   =   |\alpha_{S,D}|^{2} \frac{P_{0}}{N_0}  +  \sum \limits _{k = 1}^K C(k)|\alpha_{R_k,D}|^{2} \frac{P_{R_k}}{N_{0}}.  
 \end{equation}
It is also recalled that the MGF for   independent fading channels in DF scheme is given by \cite{SA} 
\begin{equation}\label{L12}
\text{M}_{\gamma_{\text{MRC}}}(s) = M_{\gamma_S,_D}(s)\prod\limits_{k=1}^{K}C(k) M_{\gamma R_k,_D}(s)
\end{equation}
which in the present analysis  can be  expressed according to  \cite[eq. (6)]{R7}, namely 
\begin{equation}\label{L14}
 M_{\gamma_{\eta-\mu}}\left(\frac{g}{\sin^{2}\theta}\right) = \left(\frac{4\mu^{2}h} {(2(h - H)\mu + \frac{g}{\sin^2\theta} \bar{\gamma})(2(h + H)\mu +\frac{g}  {\sin^2\theta}\bar{\gamma})}\right)^{\mu}. 
\end{equation}
It is noted  that the above expression is particularly useful in the subsequent SER analysis.

\subsection{End-to-End SER for $M{-}$PSK Constellations }
\subsubsection{The case of i.n.i.d  $\eta-\mu$ fading  channels}

The end-to-end error probability for $M{-}$PSK constellations over individual $\eta-\mu$ fading link when $ \eta$, $\mu$ and $\bar{\gamma} $ in each path are not necessarily  equal can be expressed as  \cite[eq. (5.78)]{R6}
\begin{equation}\label{L10}
  \bar{P}_{\rm M-PSK}   =  \int_{0}^{\frac{(M-1)\pi}{M}}\int_{0}^{\infty}  \frac{p_{\gamma}(\gamma)}{\pi e^{\frac{\gamma g_{\rm PSK}}{\sin^{2}\theta}}} {\rm d}\gamma {\rm d}\theta  = \underbrace{\frac{1}{\pi}\int_{0}^ {\pi/2}M_{\gamma}\left(\frac{\rm g_{PSK}}{\sin^{2}\theta}\right){\rm d}\theta}_{\triangleq \: \mathcal{I}_{1}} \; + \;\underbrace{\frac{1}{\pi} \int_{\pi/2}^ \frac{(M-1)\pi}{M}M_{\gamma}\left( \frac{g_{\rm PSK}}{\sin^{2}\theta}\right){\rm d}\theta}_{\triangleq \: \mathcal{I}_{11}} 
\end{equation}
where $ g_{\rm PSK} = \sin^{2} (\pi/M)$ \cite{R7}.  In order to evaluate  \eqref{L8}, we firstly need to determine the  error probability for decoding at the destination terminal, using  MRC,     under given decoding outcomes at nodes i.e., for a given  $\mathbf{C}_{z}$ \cite{GP}. To this end and based on the MGF approach it follows that
\begin{equation*} 
\begin{split}
 P(e|\mathbf{A} = \mathbf{C}_{z}) &= \frac{1}{\pi}\int_{0}^{\pi/2} \left(\frac{4\mu_{S,D}^{2}h_{S,D} (2(h_{S,D} + H_{S,D})\mu_{S,D} +\frac{g_{\rm PSK}}{\sin^2\theta}\bar{\gamma}_{S,D})^{-1}} {\,  (2(h_{S,D} - H_{S,D})\mu_{S,D} + \frac{g_{\rm PSK}}{\sin^2\theta} \bar{\gamma}_{S,D})}\right)^{\mu_{S,D}}  \\
&  \times \prod_{k = 1}^K C(k) \left(\frac{4\mu_{R_k,D}^{2}h_{R_k,D} \,  (2(h_{R_k,D} - H_{R_k,D})\mu_{R_k,D} + \frac{g_{\rm PSK}}{\sin^2\theta} \bar{\gamma}_{R_k,D})^{-1} } {2(h_{R_k,D} + H_{R_k,D})\mu_{R_k,D} +\frac{g_{\rm PSK}}{\sin^2\theta}\bar{\gamma}_{R_k,D}}\right)^{\mu_{R_k,D}} {\rm d} \theta
 \end{split}
\end{equation*}
\begin{equation}   \label{L15}
\begin{split}
 \qquad  \,  \qquad     &+ \frac{1}{\pi}\int_{\pi/2}^\frac{(M-1)\pi}{M} \left(\frac{4\mu_{S,D}^{2}h_{S,D}(2(h_{S,D} + H_{S,D})\mu_{S,D} +\frac{g_{\rm PSK}}{\sin^2\theta}\bar{\gamma}_{S,D})^{-1}} {(2(h_{S,D} - H_{S,D})\mu_{S,D} + \frac{g_{\rm PSK}}{\sin^2\theta} \bar{\gamma}_{S,D})}\right)^{\mu_{S,D}} \\
&  \times \prod_{k = 1}^K C(k) \left(\frac{4\mu_{R_k,D}^{2}h_{R_k,D} \,  (2(h_{R_k,D} - H_{R_k,D})\mu_{R_k,D} + \frac{g_{\rm PSK}}{\sin^2\theta} \bar{\gamma}_{R_k,D})^{-1} } {(2(h_{R_k,D} + H_{R_k,D})\mu_{R_k,D} +\frac{g_{\rm PSK}}{\sin^2\theta}\bar{\gamma}_{R_k,D})}\right)^{\mu_{R_k,D}} {\rm d}\theta. \end{split}
\end{equation}
Evidently, the derivation of an analytic solution for  \eqref{L15} is subject to analytic evaluation of the   integrals $\mathcal{I}_{1} $ and $ \mathcal{I}_{11}$ in closed-form. To this end,    for the case of non-identical fading parameters,   i.e, $ \mu_{S,D} \neq  \mu_{R_1,D} \neq \cdots  \neq  \mu_{R_K,D} $, $ \eta_{S,D} \neq  \eta_{R_1,D} \neq \cdots  \neq  \eta_{R_K,D} $ and $ \overline{\gamma}_{S,D} \neq  \overline{\gamma}_{R_1,D}  \neq \cdots \neq  \overline{\gamma}_{R_k,D} $,  the $\mathcal{I}_{1}$ term can be alternatively expressed as follows
\begin{equation} \label{LL15}
 \mathcal{I}_{1} = \frac{1}{\pi}\int_{0}^{\pi/2} \frac{1} {\left(1 + \frac{A_{1}}{\sin^2\theta}\right)^{\mu_{S,D}}\left(1 +\frac{A_{2}}{\sin^2\theta}\right)^{\mu_{S,D}}}\prod_{k = 1}^{K}\frac{C(k)  \,  {\rm d}\theta  }{\left(1 + \frac{B_{1k}}{\sin^2\theta}\right)^{\mu_{R_k,D}}\left(1 +\frac{B_{2k}}{\sin^2\theta}\right)^{\mu_{R_k,D}}}  
\end{equation} 
where
\begin{equation}
\big\{ ^{A_1}_{A_2} \big\}  = \frac{\overline{\gamma}_{S, D} g_{\rm PSK}}{2(h_{S,D} \, \{ \mp \} \, H_{S,D})\mu_{S, D}}
\end{equation}
and 
\begin{equation}
\big\{^{B_{1k}}_{B_{2k}} \big\} = \frac{\overline{\gamma}_{R_{k}, D} g_{\rm PSK}}{2(h_{R_{k},D} \{ \mp \} H_{R_{k},D})\mu_{R_{k}, D}}. 
\end{equation}
 By also setting $u = \sin^{2}(\theta)$ and carrying out tedious but basic algebraic manipulations yields
\begin{equation}\label{LL17}
\mathcal{I}_{1} = \frac{\beta_{\gamma_{\rm MRC}}(g_{\rm PSK}) }{2\pi}\int_{0}^{1}\frac{(1-u)^{-\frac{1}{2}}u^{2(\mu_{S,D}+\sum_{k = 1}^{K}C(k) \mu_{R_k,D})-\frac{1}{2}}}{\left(1 +\frac{u}{A_{1}}\right)^{\mu_{S,D}}\left(1 +\frac{u}{A_{2}}\right)^{\mu_{S,D}}}\prod_{k=1}^{K}\frac{{C(k) \, \rm d}u}{\left(1 +\frac{u}{B_{1k}}\right)^{\mu_{R_k,D}}\left(1 +\frac{u}{B_{2k}}\right)^{\mu_{R_k,D}}}
\end{equation} 
where 
\begin{equation}
\beta_{\gamma_{\rm MRC}}(g_{\rm PSK}) = \left(\frac{4\mu_{S,D}^{2} (h^{2}_{S,D} - H^{2}_{S,D})}{\overline{\gamma}_{S,D}^{2}g^{2}_{\rm PSK}}\right)^{\mu_{S,D}} \prod_{k = 1}^{K} C(k)\left(\frac{4\mu_{R_k,D}^{2}(h^{2}_{R_k, D} - H^{2}_{R_k, D})}{\overline{\gamma}_{R_k,D}^{2}g^{2}_{\rm PSK}}\right)^{\mu_{R_k,D}}.
\end{equation} 
Importantly, eq.  \eqref{LL17} can be expressed in closed-form in terms  of \cite[eq. (7.2.4.57)]{pud}, yielding   
\begin{equation}\label{LLL20}
\mathcal{I}_{1}  =  \frac{\beta_{\gamma_{\rm MRC}}(g_{\rm PSK})\Gamma(2\mu_{S,D}+ 2\sum_{k = 1}^{K}C(k)\mu_{R_k,D} + \frac{1}{2}) }{2\sqrt{\pi}\Gamma(2\mu_{S,D} + 2\sum_{k = 1}^{K}C(k)\mu_{R_k,D} + 1)}   \hspace{7cm}
\end{equation} 
\begin{equation*}
\begin{split}
  \times  F_{D}^{(2K +2)}\left(2\mu_{S,D} + 2\sum_{k = 1}^{K}C(k)\mu_{R_k,D} +\frac{1}{2} ;\mu_{S,D},\mu_{S,D}, \mu_{R_1,D}, \cdots, \mu_{R_{K},D}, \mu_{R_1,D}, \cdots, \mu_{R_{K},D} \right. \hspace{1.1cm} \\  
\, \hspace{1.1cm} \left.  ;2\mu_{S,D} + 2\sum_{k = 1}^{K}C(k)\mu_{R_k,D} + 1; -\frac{ 1}{A_{1}}, -\frac{1}{A_{2}}, -\frac{1}{B_{11}},\cdots,-\frac{ 1}{B_{1K}}, -\frac{1}{B_{21}},\cdots,-\frac{1}{B_{2,K}}\right) 
\end{split}
\end{equation*} 
where $F^{(n)}_{D}(\cdot) $ denotes the generalized Lauricella   hypergeometric function of $n$  variables  \cite{NYY}.
 
In the same context, for the $\mathcal{I}_{11}$ integral we set $ u = \cos^2(\theta)/\cos^2(\pi/M)$ in \eqref{L15} yielding
\begin{equation}\label{GG}
\mathcal{I}_{11} =  \frac{M_{\gamma_{\rm MRC}}(g_{\rm PSK}) \cos(\pi/M)}{2\pi}   \hspace{10cm}
\end{equation} 
\begin{equation*}
\hspace{1cm} \times \int_{0}^{1}\frac{u^{-\frac{1}{2}}(1-\cos^{2}(\pi/M)u)^{{2(\mu_{S,D}+\sum_{k = 1}^{K}C(k) \mu_{R_k,D})-\frac{1}{2}}}}{\left(1 -\frac{\cos^{2}(\pi/M)}{1 + A_{1}}u\right)^{\mu_{S,D}}\left(1 -\frac{\cos^{2}(\pi/M)}{1+A_{2}}u\right)^{\mu_{S,D}}}\prod_{k=1}^{K}C(k)\frac{ \left(1 -\frac{\cos^{2}(\pi/M)}{1+B_{2k}}u\right)^{-\mu_{R_k,D}} }{\left(1 -\frac{\cos^{2}(\pi/M)}{1+B_{1k}}u\right)^{\mu_{R_k,D}}} {\rm d}u.
\end{equation*}
Evidently, the  above integral can be also expressed in closed-form in terms of the  $F^{(n)}_{D}(\cdot) $ function; therefore, by performing the necessary change of variables and substituting in \eqref{GG}, one obtains
\begin{equation} \label{LLLL20}
\mathcal{I}_{11} =  \frac{M_{\gamma_{\rm MRC}}(g_{\rm PSK}) }{\pi}   \hspace{13.2cm}
\end{equation}
 \begin{equation*}
\begin{split}
& \times  F_{D}^{(2K +3)}\left(\frac{1}{2} ;\mu_{S,D},\mu_{S,D}, \mu_{R_1,D},\cdots, \mu_{R_K,D}, \mu_{R_{1},D},\cdots,\mu_{R_K,D}, \frac{1}{2}-2\mu_{S,D} + 2\sum_{k = 1}^{K}C(k)\mu_{R_K,D};\frac{3}{2};\right.  \hspace{17.25cm}\\  
&  \left.\frac{\cos^{2}(\pi/M)}{1+A_{1}}, \frac{\cos^{2}(\pi/M)}{1+A_{2}}, \frac{\cos^{2}(\pi/M)}{1+B_{11}},\cdots,\frac{\cos^{2}(\pi/M)}{1+B_{1K}}, \frac{\cos^{2}(\pi/M)}{1+B_{21}},\cdots,\frac{\cos^{2}(\pi/M)}{1+B_{2K}}, \cos^{2}(\pi/M) \right). \hspace{1.75cm}
\end{split}
\end{equation*}   

It is noted here that the $F^{(n)}_{D}(\cdot) $  function   has been studied extensively over the past decades. Nevertheless, despite its importance it is not unfortunately included as built-in function in popular  software packages such as MATLAB, MATHEMATICA and MAPLE. Based on this, a simple MATLAB algorithm for computing this function straightforwardly is proposed in Appendix I. 
\subsubsection{The  case of i.i.d $\eta{-}\mu $ fading channels}
For the special case of identical relay to destination paths i.e.,  $ \mu_{R_1,D}  =   \cdots  = \mu_{R_K,D} =  \mu_{R,D}, \eta_{R_1,D}  =  \cdots  = \eta_{R_K,D} = \eta_{R,D}$, $\overline{\gamma}_{R_1,D}  =  \cdots =  \overline{\gamma}_{R_K,D}  = \overline{\gamma}_{R,D} $ and thus, $ B_{11} = \cdots = B_{1K} = B_{1} $,  and  $ B_{21} = \cdots = B_{2K} = B_{2} $, equations  \eqref{LLL20} and \eqref{LLLL20} can be readily  simplified to the following representations
\begin{equation}\label{III1}
\begin{split}
\mathcal{I}_{1 {\rm i.i.d}} =  \frac{\beta_{\gamma_{\rm MRC}}(g_{\rm PSK}) \Gamma\left(2\mu_{S,D}+  2\mu_{R,D}\sum_{k = 1}^{K} C(k)  +\frac{1}{2}\right)}{2\sqrt{\pi}\Gamma\left(2\mu_{S,D}+ 2\mu_{R,D} \sum_{k = 1}^{K}C(k)+1\right)} 
 F_{D}^{(4)}\left(2\mu_{S,D}+  2\mu_{R,D}\sum_{k = 1}^{K} C(k)  +\frac{1}{2}; \right. \hspace{2.0cm} \\   
\left.  \mu_{S,D}, \mu_{S,D}, K\mu_{R,D},  K\mu_{R,D}; 2\mu_{S,D}+  2\mu_{R,D}\sum_{k = 1}^{K} C(k)  + 1; -\frac{1}{A_{1}}, -\frac{1}{A_{2}}, -\frac{1}{B_{1}},-\frac{1}{B_{2}}\right)  \hspace{1.5cm} 
\end{split}
\end{equation} 
\text{and}
\begin{equation}\label{III2}
\begin{split}
\mathcal{I}_{11 {\rm i.i.d}} =  \frac{M_{\gamma_{\rm MRC}}(g_{\rm PSK}) }{\pi} 
 F_{D}^{(5)}\left(\frac{1}{2} ;\mu_{S,D},\mu_{S,D},   K\mu_{R,D},   K\mu_{R,D},\frac{1}{2} - 2\mu_{S,D} - 2\mu_{R,D} \sum_{k = 1}^{K} C(k) \right.  \hspace{-0.5cm}\\  
\left.;\frac{3}{2};\frac{\cos^{2}(\pi/M)}{1+A_{1}}, \frac{\cos^{2}(\pi/M)}{1+A_{2}}, \frac{\cos^{2}(\pi/M)}{1+B_{1}},\frac{\cos^{2}(\pi/M)}{1+B_{2}}, \cos^{2}(\pi/M) \right).
\end{split}
\end{equation}
respectively. As a result, with the aid of the derived   expressions for $\mathcal{I}_{1} $ and $\mathcal{I}_{11} $, 
the corresponding error probability  for $M{-}$PSK constellations  can be   determined by  $P(e|\mathbf{A} = \mathbf{C}_{z}) = \mathcal{I}_{1} + \mathcal{I}_{11}.$  

It is recalled that the derivation of the overall SER also requires the determination of the decoding probability of the relay nodes $ P(\mathbf{A}= \mathbf{C}_{z} ) $. This is in fact a direct product of the terms $ P(\bar{\gamma}_{S,R_k}) =  P(A(k) = C(k) = 0 )$ i.e. decoding error at the relays and $ (1- P(\bar{\gamma}_{S,R_k})) =  P(A(k) = C(k) = 1 )$ i.e. successful decoding at the relays, which, as already mentioned, is a pre-requisite for forwarding re-encoded signals to the destination. Importantly, this can be also determined in closed-form with the aid of the commonly used MGF approach, namely
\begin{equation}\label{SS}
P(\bar{\gamma}_{S,R_k}) = \underbrace{\frac{1}{\pi}\int_{0}^ {\pi/2}M_{\gamma_{S,R_k}}\left(\frac{g_{\rm PSK}}{\sin^{2}\theta}\right) {\rm d}\theta}_{\triangleq \: \mathcal{I}_{2}} \; + \;\underbrace{\frac{1}{\pi} \int_{\pi/2}^ \frac{(M-1)\pi}{M}M_{\gamma_{S,R_k}}\left( \frac{g_{\rm PSK}}{\sin^{2}\theta}\right){\rm d}\theta}_{\triangleq \: \mathcal{I}_{22}}
\end{equation}
which with the aid of \eqref{L14} can be equivalently re-written as follows
\begin{equation*}
P(A(k) = C(k) = 0)  = \frac{1}{\pi}\int_{0}^{\pi/2} \left(\frac{4\mu_{S,R_k}^{2}h_{S,R_k} \, (2(h_{S,R_k} + H_{S,R_k})\mu_{S,R_k} + \frac{g_{\rm PSK}}{\sin^2\theta}\bar{\gamma}_{S,R_k})^{-1}} {(2(h_{S,R_k} - H_{S,R_k})\mu_{S,R_k} + \frac{g_{\rm PSK}}{\sin^2\theta} \bar{\gamma}_{S,R_k})}\right)^{\mu_{S,R_k}} {\rm d}\theta  
\end{equation*}
\begin{equation} \label{KK}
 + \frac{1}{\pi}\int_{\pi/2}^\frac{(M-1)\pi}{M} \left(\frac{4\mu_{S,R_k}^{2}h_{S,R_k} \, (2(h_{S,R_k} + H_{S,R_k})\mu_{S,R_k} + \frac{g_{\rm PSK}}{\sin^2\theta}\bar{\gamma}_{S,R_k})^{-1} } {(2(h_{S,R_k} - H_{S,R_k})\mu_{S,R_k} + \frac{g_{\rm PSK}}{\sin^2\theta} \bar{\gamma}_{S,R_k})}\right)^{\mu_{S,R_k}} {\rm d}\theta. 
\end{equation}
Notably,  the integrals in \eqref{KK} have  similar algebraic form to  $\mathcal{I}_{1} $ and $\mathcal{I}_{11}$ since the difference is the absence of  $ \mu_{R_k,D} $ terms.  As a result, the following closed-form expressions can be   deduced 
\begin{equation}\label{LLL21}
\mathcal{I}_{2} =  \frac{\beta_{\gamma_{S,R_k}}(g_{\rm PSK}) \Gamma(2\mu_{S,R_k}+\frac{1}{2})}{2\sqrt{\pi}\Gamma(2\mu_{S,R_k}+1)} 
 F_{D}^{(2)}\left(2\mu_{S,R_k}+\frac{1}{2};\right. \left. 
 \mu_{S,R_k},\mu_{S,R_k}; 2\mu_{S,R_k}+ 1; -\frac{1}{C_{1}}, -\frac{1}{C_{2}}\right) 
\end{equation} 
and
\begin{equation}\label{LLI22}
\mathcal{I}_{22} =  \frac{M_{\gamma_{S,R_k}}(g_{\rm PSK}) }{\pi} 
 F_{D}^{(3)}\left(\frac{1}{2} ;\mu_{S,R_k},\mu_{S,R_k},\frac{1}{2}-2\mu_{S,R_k}  
;\frac{3}{2};\frac{\cos^{2}(\pi/M)}{1+C_{1}}, \frac{\cos^{2}(\pi/M)}{1+C_{2}}, \cos^{2}(\pi/M)\right) 
\end{equation} 
where 
\begin{equation} \label{statheres}
\big\{^{C_{1}}_{C_{2}} \big\} =  \frac{\overline{\gamma}_{S, R_{k}} g_{\rm PSK}}{2(h_{S, R_{k}} \{ \mp \} H_{S,R_{k}})} 
\end{equation} 
and
\begin{equation}
\beta_{S,R_k}(g_{\rm PSK}) = \left(\frac{4\mu_{S,R_k}^{2}(h^{2}_{S,R_k} - H^{2}_{S,R_k})}{\overline{\gamma}_{S,R_k}^{2}g^{2}_{\rm PSK}}\right)^{\mu_{S,R_k}}.\end{equation}
Therefore, the $P_{\rm SER}^{D}$ for $M-$PSK is deduced by  inserting $P(e|\mathbf{A} = \mathbf{C}_{z})$ and $P(\mathbf{A} = \mathbf{C}_{z})  $ in \eqref{L8}. 

To the best of the authors' knowledge, the derived  analytic expressions are novel. 

\subsection{ End-to-End SER   for $M{-}$QAM Constellations}

Having derived the SER over i.n.i.d and i.i.d $\eta-\mu$ fading channels for the case of $M-$PSK modulations, this subsection is devoted to the derivation of  the corresponding  SER  for the case of  $M{-}$QAM constellations. 

\subsubsection{The case of i.n.i.d $\eta{-}\mu$ fading channels}

In the    case of  independent but not necessarily identically distributed $\eta-\mu$ fading channels and based on  \cite[eq.  (9.21)]{R6}, it follows that 
\begin{equation}\label{LL32}  
\bar{P}_{\rm M-QAM} = \frac{4C}{\pi}\underbrace{\int_{0}^ {\pi/2}M_{\gamma_{\rm MRC}}\left(\frac{g_{\rm QAM}}{\sin^{2}\theta}\right) {\rm d}\theta}_{\triangleq \: \mathcal{I}_{3}} -\frac{4C^{2}}{\pi}\underbrace{\int_{0}^ {\pi/4}M_{\gamma_{\rm MRC}}\left( \frac{g_{\rm QAM}}{\sin^{2}\theta}\right) {\rm d}\theta}_{\triangleq \: \mathcal{I}_{4}}
\end{equation}
where $ g_{\rm QAM} = 3/2(M-1)$ and $C = (1 - 1/\sqrt{M})$ \cite{GP}. Therefore, the probability of decoding error using  MRC  under given decoding results at the relays can be expressed as follows:
\begin{equation} \label{LLL33}
 P(e|\mathbf{A} = \mathbf{C}_{z}) = \frac{4C}{\pi}\int_{0}^{\pi/2} \left(\frac{4\mu_{S,D}^{2}h_{S,D}(2(h_{S,D} + H_{S,D})\mu_{S,D} +\frac{g_{\rm QAM}}{\sin^2\theta}\bar{\gamma}_{S,D})^{-1} } {(2(h_{S,D} - H_{S,D})\mu_{S,D} + \frac{g_{\rm QAM}}{\sin^2\theta} \bar{\gamma}_{S,D})}\right)^{\mu_{S,D}}  
 \end{equation}
 \begin{equation*}
  \qquad \qquad \qquad \qquad \times\prod_{k = 1}^K C(k) \left(\frac{4\mu_{R_k,D}^{2}h_{R,_k,D} \, (2(h_{R_k,D} + H_{R_k,D})\mu_{R_k,D} +\frac{g_{\rm QAM}}{\sin^2\theta}\bar{\gamma}_{R_k,D})^{-1}} {(2(h_{R_k,D} - H_{R_k,D})\mu_{R_k,D} + \frac{g_{\rm QAM}}{\sin^2\theta} \bar{\gamma}_{R_k,D})}\right)^{\mu_{R_k,D}} {\rm d}\theta 
 \end{equation*}
\begin{equation*}  
\begin{split}
 &  \qquad \qquad \qquad \quad \; - \frac{4C^{2}}{\pi}\int_{0}^{\pi/4} \left(\frac{4\mu_{S,D}^{2}h_{S,D} (2(h_{S,D} + H_{S,D})\mu_{S,D} +\frac{g_{\rm QAM}}{\sin^2\theta}\bar{\gamma}_{S,D})^{-1}} {(2(h_{S,D} - H_{S,D})\mu_{S,D} + \frac{g_{\rm QAM}}{\sin^2\theta} \bar{\gamma}_{S,D})}\right)^{\mu_{S,D}} \\
&   \qquad \qquad \qquad \quad \; \times \prod_{k = 1}^K C(k) \left(\frac{4\mu_{R_k,D}^{2}h_{R_k,D} \, (2(h_{R_k,D} + H_{R_k,D})\mu_{R_k,D} +\frac{g_{\rm QAM}}{\sin^2\theta}\bar{\gamma}_{R_k,D})^{-1} } {(2(h_{R_k,D} - H_{R_k,D})\mu_{R_k,D} + \frac{g_{\rm QAM}}{\sin^2\theta} \bar{\gamma}_{R_k,D})}\right)^{\mu_{R_k,D}} {\rm d}\theta.
\end{split}
\end{equation*}
Evidently, the derivation of a closed-form expression for  \eqref{LLL33} is subject to analytic evaluation of the above two integrals that correspond to $ \mathcal{I}_{3}$  and $ \mathcal{I}_{4} $. Following the same  methodology  as in the case of   $M{-}$PSK constellations and using the change of variable of $ u = \sin^{2} \theta $, one obtains
\begin{equation}\label{LLL34}
\mathcal{I}_{3} = \frac{\beta_{\gamma_{\rm MRC}}(g_{\rm QAM}) }{2}\int_{0}^{1}\frac{(1-u)^{-\frac{1}{2}}u^{2(\mu_{S,D}+\sum_{k = 1}^{K}C(k) \mu_{R_k,D})-\frac{1}{2}}}{\left(1 +\frac{u}{A_{1}}\right)^{\mu_{S,D}}\left(1 +\frac{u}{A_{2}}\right)^{\mu_{S,D}}}\prod_{k=1}^{K}\frac{C(k) \, {\rm d}u}{\left(1 +\frac{u}{B_{1k}}\right)^{\mu_{R_k,D}}\left(1 +\frac{u}{B_{2k}}\right)^{\mu_{R_k,D}}}
\end{equation} 
where the parameters $A_{1},A_{2}, B_{1k} \;\text{and}\; B_{2k} $   are determined by substituting  $g_{\rm PSK}$ with $g_{\rm QAM}$. To this effect and after some algebraic manipulations, the following  analytic expression is deduced  
\begin{equation}\label{LLL35}
\mathcal{I}_{3} =  \frac{\beta_{\gamma_{\rm MRC}}(g_{\rm QAM})\sqrt{\pi} \Gamma\left(2\mu_{S,D} + 2\sum_{k = 1}^{K}C(k)\mu_{R_k,D}+\frac{1}{2} \right) }{2\Gamma\left(2\mu_{S,D}+ 2\sum_{k = 1}^{K}C(k)\mu_{R_k,D}+ 1\right)} \hspace{7.8cm}
\end{equation}  
\begin{equation*} 
\begin{split}
&  \times  F_{D}^{(2K +2)}\left(2\mu_{S,D} + 2\sum_{k = 1}^{K}C(k)\mu_{R_k,D}+\frac{1}{2} ;\mu_{S,D},\mu_{S,D}, \mu_{R_1,D}, \cdots, \mu_{R_{K},D}, \mu_{R_1,D}, \cdots, \mu_{R_{K},D} \right.  \qquad \\  
&   \qquad \qquad  \quad   \left. ;2\mu_{S,D} + 2\sum_{k = 1}^{K}C(k)\mu_{R_k,D}+ 1; -\frac{ 1}{A_{1}}, -\frac{1}{A_{2}}, -\frac{1}{B_{11}},\cdots,-\frac{ 1}{B_{1K}}, -\frac{1}{B_{21}},\cdots,-\frac{1}{B_{2,K}} \right).
\end{split}
\end{equation*}  
Likewise, the   $\mathcal{I}_{4}$ integral has the same integrand   but a different upper limit of integration. Thus, by following
  the same methodology and setting $y = 2u$,  yields the following   closed-form expression  
 \begin{equation} 
 \mathcal{I}_{4} =  \frac{\beta_{\gamma_{\rm MRC}}(g_{\rm QAM}) \Gamma(2\mu_{S,D}+ 2\sum_{k = 1}^{K}C(k)\mu_{R_k,D}+\frac{1}{2}) }{\Gamma(2\mu_{S,D}+ 2\sum_{k = 1}^{K}C(k)\mu_{R_k,D}+ \frac{3}{2})2^{2(\mu_{S,D}+\sum_{k =1}^{K}C(k)\mu_{R_k,D})+\frac{3}{2}}} \hspace{10cm}
  \end{equation} 
\begin{equation*} 
\begin{split}
  & \times  F_{D}^{(2K +3)}\left(2\mu_{S,D}+ 2 \sum_{k = 1}^{K}C(k)\mu_{R_k,D}+\frac{1}{2} ;\mu_{S,D},\mu_{S,D}, \mu_{R_1,D}, \cdots, \mu_{R_{K},D},\mu_{R_1,D}, \cdots, \mu_{R_{K},D}; \frac{1}{2}\right. \\
&  \quad \left.    ;2\mu_{S,D} + 2\sum_{k = 1}^{K}C(k)\mu_{R_k,D}+ \frac{3}{2}; -\frac{ 1}{2A_{1}}, -\frac{1}{2A_{2}}, -\frac{1}{2B_{11}},\cdots,-\frac{ 1}{2B_{1K}}, -\frac{1}{2B_{21}},\cdots,-\frac{1}{2B_{2K}},\frac{1}{2}\right) 
\end{split}
\end{equation*} 
where
\begin{equation}
\beta_{\gamma_{\rm MRC}}(g_{\rm QAM}) = \left(\frac{4\mu_{S,D}^{2}h_{S,D}}{\overline{\gamma}_{S,D}^{2}g^{2}_{\rm QAM}}\right)^{\mu_{S,D}} \prod_{k = 1}^{K} C(k) \left (\frac{4\mu_{R_k,D}^{2}(h^{2}_{R_k, D} - H^{2}_{R_k, D})}{\overline{\gamma}_{R_k,D}^{2}g^{2}_{\rm QAM}}\right)^{\mu_{R_k,D}}.
\end{equation} 
Notably, the $\mathcal{I}_{3}$ and $\mathcal{I}_4$ terms are also expressed in terms of the generalized Lauricella function. 
\subsubsection{The case of i.i.d $\eta{-}\mu$ fading channels}

In this simplified scenario, the corresponding solutions for   $ \mathcal{I}_{3}$  and $  \mathcal{I}_{4}$    for the case of  $M-$QAM constellations can be derived by following the same methodology as in the case of $M{-}$PSK modulations. To this end, after a necessary change of variables and long but basic algebraic manipulations, it immediately  follows that 
\begin{equation}\label{LLL38}
\begin{split}
\mathcal{I}_{3 \rm i.i.d} &=  \frac{\beta_{\gamma_{\rm MRC}}(g_{\rm QAM})\sqrt{\pi} \Gamma\left(2\mu_{S,D}+  2\mu_{R,D}\sum_{k = 1}^{K} C(k)+\frac{1}{2}\right) }{2\Gamma\left(2\mu_{S,D}+  2\mu_{R,D}\sum_{k = 1}^{K} C(k)+ 1\right)}
 F_{D}^{(4)}\left(2\mu_{S,D}+  2\mu_{R,D} \sum_{k = 1}^{K}C(k) +\frac{1}{2};   \right. \hspace{-1.25cm}\\   
& \left.  \,  \qquad \,  \mu_{S,D}, \mu_{S,D},  K\mu_{R,D},  K\mu_{R,D}; \frac{1}{2}  ;2\mu_{S,D}+   2\mu_{R,D}\sum_{k = 1}^{K} C(k)+ \frac{3}{2}; -\frac{ 1}{A_{1}}, -\frac{1}{A_{2}}, -\frac{1}{B_{1}},-\frac{ 1}{B_{2}}\right) 
\end{split}
\end{equation} 
and
\begin{equation}\label{LLL39}
\begin{split}
\mathcal{I}_{4 \rm i.i.d} &=  \frac{\beta_{\gamma_{\rm MRC}}(g_{\rm QAM}) \Gamma(2\mu_{S,D}+ 2\mu_{R,D}\sum_{k = 1}^{K} C(k)+\frac{1}{2}) }{\Gamma(2\mu_{S,D}+ 2\mu_{R,D}\sum_{k = 1}^{K} C(k) + \frac{3}{2})2^{2\mu_{S,D} + 2\mu_{R,D} \sum_{k = 1}^{K} C(k) +\frac{3}{2}}}  F_{D}^{(5)}\left(2\mu_{S,D}+ 2\mu_{R,D}\sum_{k = 1}^{K} C(k) +\frac{1}{2};\right. \\   
& \left. \mu_{S,D}, \mu_{S,D},   K\mu_{R,D},  K\mu_{R,D}, \frac{1}{2};2\mu_{S,D}+ 2\mu_{R_k,D}\sum_{k = 1}^{K} C(k)+ \frac{3}{2}; -\frac{ 1}{2A_{1}}, -\frac{1}{2A_{2}}, -\frac{1}{2B_{1}},-\frac{ 1}{2B_{2}},\frac{1}{2} \right).
\end{split}
\end{equation} 
respectively. Hence,  the  error probability is determined  by $ P(e|\mathbf{A} = \mathbf{C}_{z}) = 4C \mathcal{I}_{3}/\pi - 4C^{2}\mathcal{I}_{4}/\pi$.  

Likewise, the corresponding  decoding error probability at the relay nodes    can be expressed as
\begin{equation}\label{LLL40}
P(\bar{\gamma}_{S,R_k}) = \frac{4C}{\pi}\underbrace{\int_{0}^ {\pi/2}M_{\gamma_{S,R_k}}\left(\frac{g_{\rm QAM}}{\sin^{2}\theta}\right) {\rm d} \theta}_{\triangleq \: \mathcal{I}_{5}} -\frac{4C^{2}}{\pi}\underbrace{\int_{0}^ {\pi/4}M_{\gamma_{S,R_k}}\left( \frac{g_{\rm QAM}}{\sin^{2}\theta}\right) {\rm d}\theta.}_{\triangleq \: \mathcal{I}_{6}}
\end{equation} 
Based on the approach   in \eqref{KK} and given  the non-arbitrary limits of integration, the following exact closed-form expressions are deduced for the integrals  $\mathcal{I}_{5}$ and $ \mathcal{I}_{6}$:  
\begin{equation}\label{LLL42}
\begin{split}
\mathcal{I}_{5} =&  \frac{\beta_{\gamma_{S,R_k}}(g_{\rm QAM})\sqrt{\pi} \Gamma(2\mu_{S,R_k} \sum_{k = 1}^{K} C(k) +\frac{1}{2}) }{2\Gamma(2\mu_{S,R_k} \sum_{k = 1}^{K} C(k) + 1)}  \\
& \times  F_{D}^{(2)}\left(2\mu_{S,R_k}\sum_{k = 1}^{K} C(k) +\frac{1}{2} ;\mu_{S,R_k},\mu_{S,R_k}  
 ;2\mu_{S,R_k}+ 1; -\frac{ 1}{C_{1}}, -\frac{1}{C_{2}}\right) 
 \end{split}
\end{equation} 
and 
\begin{equation}\label{LLL43}
\begin{split}
\mathcal{I}_{6} =&  \frac{\beta_{\gamma_{S,R_k}}(g_{\rm QAM}) \Gamma(2\mu_{S,R_k}\sum_{k = 1}^{K} C(k) +\frac{1}{2}) }{\Gamma(2\mu_{S,R_k} \sum_{k = 1}^{K} C(k) +\frac{3}{2})2^{2\mu_{S,R_k} \sum_{k = 1}^{K} C(k) +\frac{3}{2}}}  \quad \\
&  \times F_{D}^{(3)}\left(2\mu_{S,R_k} \sum_{k = 1}^{K} C(k) +\frac{1}{2} ;\mu_{S,R_k},\mu_{S,R_k}, \frac{1}{2}
 ;2\mu_{S,R_k} \sum_{k = 1}^{K} C(k) + \frac{3}{2}; -\frac{ 1}{2C_{1}}, -\frac{1}{2C_{2}},\frac{1}{2}\right) 
 \end{split}
\end{equation}    
respectively, where
\begin{equation}
  \beta_{\gamma_{S,R_k}}(g_{\rm QAM}) = \left( \frac{4\mu_{S,R_k}^{2}(h^{2}_{S,R_k}  -  H^{2}_{S,R_k})}{\overline{\gamma}_{S,R_k}^{2}g^{2}_{\rm QAM}} \right)^{\mu_{S,R_k}}
\end{equation}
 whereas   $C_{1} \;\text{and}\; C_{2} $  are obtained by substituting  $g_{\rm PSK}$ with $g_{\rm QAM}$ in \eqref{statheres}.
Therefore, using \eqref{LLL42} and \eqref{LLL43}, the corresponding  decoding error probability can be readily expressed as 
\begin{equation}
P(\bar{\gamma}_{S,R_k}) = P(A(k) = C(k) = 0) = \frac{4C}{\pi} \mathcal{I}_{5} - \frac{4C^{2}}{\pi} \mathcal{I}_{6}. 
\end{equation}
Thus,  a closed-form expression for the   SER is deduced by substituting $P(\mathbf{A}= \mathbf{C}_{z})$ in  \eqref{L8} along with the corresponding  $P(e|\mathbf{A} = \mathbf{C}_{z}) $ and can be computed using  the  algorithm in Appendix I. 

\subsection{Simple Asymptotic Expressions}

The derivation of asymptotic expressions typically leads to useful  insights on the impact of the involved parameters on the system performance. This is also the case in the present analysis as simple closed-form asymptotic  expressions are derived for high SNR values. To this end,  the MGF of  $\eta-\mu$ distribution can be accurately approximated as  
\begin{equation} \label{LLL44}
 M_{\gamma_{\eta-\mu}}\left(\frac{g}{\sin^{2}\theta}\right) = \left(\frac{4\mu^{2}h} {(2(h - H)\mu + \frac{g}{\sin^2\theta} \bar{\gamma})(2(h + H)\mu +\frac{g}  {\sin^2\theta}\bar{\gamma})}\right)^{\mu} \approx  \left(\frac{4\mu^{2}h}{g^2\bar{\gamma}^2}\right)^{\mu}\sin^{4\mu}(\theta). 
\end{equation}
Based on this,  the conditional error probability $P(e|\mathbf{A} = \mathbf{C}_{z})$   can be approximated as  follows: 
\begin{equation}\label{LLL45}
P(e|\mathbf{A} = \mathbf{C}_{z}) \approx \left(\frac{4\mu_{S,D}^{2}h_{S,D}}{g^{2}\overline{\gamma}_{S,D}^{2}}\right)^{\mu_{S,D}}\: \prod_{k=1}^{K}\;A_{R_k,D}(\mathbf{C}_{z})\left(\frac{4\mu_{R,D}^{2}h_{R_k,D}}{g^{2}\overline{\gamma}_{R_k,D}^{2}}\right)^{\mu_{R_k,D}}
\end{equation}
where $ A_{R_k,D}(\mathbf{C}_{z}) $ for $M{-}$PSK constellations is given by 
\begin{equation} \label{MM1}
A_{R_k,D}(\mathbf{C}_{z}) = \frac{1}{\pi} \int_{0}^{\frac{(M-1)\pi}{\pi}} [\sin(\theta)]^{4(\mu_{S,D} + \sum_{k=1}^{K}C(k)\mu_{R_k,D})} {\rm d}\theta    
 \end{equation}
which can be equivalently re-written as follows: 
\begin{equation}
A_{R_k,D}(\mathbf{C}_{z})  = \frac{1}{\pi} \underbrace{\int_{0}^{\pi/2} [\sin(\theta)]^{4(\mu_{S,D} + \sum_{k=1}^{K}C(k)\mu_{R_k,D})}{\rm d}\theta}_{ \mathcal{I}_{7}} + \frac{1}{\pi} \underbrace{\int_{\pi/2}^{\frac{(M-1)\pi}{M}} [\sin(\theta)]^{4(\mu_{S,D} + \sum_{k=1}^{K}C(k)\mu_{R_k,D})}{\rm d}\theta}_ {\mathcal{I}_{8}}.
\end{equation}
The  derivation of a closed-form expression for \eqref{LLL45} is subject to analytic evaluation of the above   trigonometric integrals. Hence,   setting  $ u = \sin^{2}(\theta) $ and  $ a = 2(\mu_{S,D} + \sum_{k=1}^{K}C(k)\mu_{R_k,D}) $ yields
\begin{equation}\label{MM2}
\mathcal{I}_{7} = \frac{1}{2}\int_{0}^{1}\frac{u^{a-\frac{1}{2}}}{(1-u)^{\frac{1}{2}}}{\rm d}u
\end{equation}
which can be expressed in closed-form according to \cite[eq. (3.191.1)]{GR}. To this end, by performing the necessary variable transformation and after   basic algebraic manipulations one obtains
\begin{equation}\label{MM3}
\mathcal{I}_{7} = 
\frac{\sqrt{\pi}\Gamma(2\mu_{S,D} + 2\sum_{k=1}^{K}C(k)\mu_{R_k,D}+\frac{1}{2})}{2\Gamma(2\mu_{S,D} + 2\sum_{k=1}^{K}C(k)\mu_{R_k,D}+1)}.
\end{equation}
Likewise, by setting $ u = \cos^{2}(\theta)/\cos ^{2}(\pi/M)$ to the second integral in  \eqref{MM1}, it follows that  
\begin{equation}\label{MM5}
\mathcal{I}_{8} = \frac{\cos(\pi/M)}{2}\int_{0}^{1}\frac{u^{-\frac{1}{2}}}{\left(1-u\cos^2\left( \frac{\pi}{M} \right) \right)^{\frac{1}{2}- a}} {\rm d}u
\end{equation}  
which can be expressed in closed-form in terms of the Gauss hypergeometric function, yielding  
\begin{equation}\label{MM6}
\mathcal{I}_{8} = \cos \left(\frac{\pi}{M} \right)  \;_{2}F_{1} \left( \frac{1}{2}-2\mu_{S,D} + 2\sum_{k=1}^{K}C(k)\mu_{R_k,D},\frac{1}{2};  \frac{3}{2}; \cos^2 \left( \frac{\pi}{M}  \right) \right). 
\end{equation}
To this effect,  $ A_{R_k,D}(\mathbf{C}_{z}) $ can be expressed in  closed-form by
$  A_{R_k,D}(\mathbf{C}_{z}) = \mathcal{I}_{7}/\pi  + \mathcal{I}_{8}/\pi.$ 

In the same context, for the case of   M${-}$QAM  constellations one obtains
\begin{equation}\label{MK}
A_{R_k,D}(\mathbf{C}_{z}) = \frac{4C}{\pi} \underbrace{\int_{0}^{\pi/2} [\sin(\theta)]^{4(\mu_{S,D} + \sum_{k=1}^{K}C(k)\mu_{R_k,D})} {\rm d}\theta}_{\mathcal{I}_{9}} \; - \; \frac{4C^2}{\pi} \underbrace{ \int_{0}^{\pi/4} [\sin(\theta)]^{4(\mu_{S,D} + \sum_{k=1}^{K}C(k)\mu_{R_k,D})} {\rm d}\theta}_{\mathcal{I}_{10}}. 
\end{equation}   
It is evident that the integrals $\mathcal{I}_7$ and $\mathcal{I}_9$  have the same algebraic forms.  Hence, it   follows that
\begin{equation}\label{MM7}
\mathcal{I}_{9} = 
\frac{\sqrt{\pi}\Gamma(2(\mu_{S,D} + \sum_{k=1}^{K}C(k)\mu_{R_k,D})+\frac{1}{2})}{2\Gamma(2(\mu_{S,D} + \sum_{k=1}^{K}C(k)\mu_{R_k,D})+1)}
\end{equation}
for the first integral, while and upon setting $ u = 2\sin^{2}(\theta) $ in the second integral, one obtains  
\begin{equation}\label{MM8}
\mathcal{I}_{10} = 
\frac{\, 2^{-(2\mu_{S,D}+\frac{3}{2})}\,_{2}F_{1} \left(\frac{1}{2},2\mu_{S,D} + 2\sum_{k=1}^{K}C(k)\mu_{R_k,D}+\frac{1}{2}; 2\mu_{S,D} + 2\sum_{k=1}^{K}C(k)\mu_{R_k,D}+\frac{3}{2}; \frac{1}{2}\right)}{2^{2\sum_{k=1}^{K}C(k)\mu_{R_k,D}}\;\Gamma(2\mu_{S,D} + 2\sum_{k=1}^{K}C(k)\mu_{R_k,D}+\frac{3}{2}) [\Gamma(2\mu_{S,D} + 2\sum_{k=1}^{K}C(k)\mu_{R_k,D} +\frac{1}{2})]^{-1}}. 
\end{equation} 
Therefore, with the aid of  \eqref{MM7} and \eqref{MM8}, a closed-form expression for   $ A_{R_k,D}(\mathbf{C}_{z}) $ for the case of  square $M{-}$QAM constellations   is given by $  A_{R_k,D}(\mathbf{C}_{z}) = 4C\mathcal{I}_{9}/\pi  - 4C^2\mathcal{I}_{10}/\pi$. 

The decoding error probability of the relays  also in the  high SNR regime can be expressed as   
\begin{equation}\label{X}
P(A(k) = C(k) = 0) \approx A_{S,R_k}\left(\frac{4\mu_{S,R_k}^{2}h_{S,R_k}}{g^{2}\overline{\gamma}_{S,R_k}^{2}}\right)^{\mu_{S,R_k}}
\end{equation}  
and
\begin{equation}  
P(A(k)= C(k) = 1) \approx 1 - A_{S,R_k} \left(\frac{4\mu_{S,R_k}^{2}h_{S,R_k}}{g^{2} \overline{\gamma}_{S,R_k}^{2}}\right)^{\mu_{S,R_k}}
\end{equation}  
where $ A_{S,R_k}$ for $M{-}$PSK and $M{-}$QAM constellations is given by 
\begin{equation}\label{MK2}
 A_{S,R_k} = \frac{1}{\pi} \int_{0}^\frac{(M-1)\pi}{M} \sin^{4\mu_{S,R_k}}(\theta) {\rm d}\theta
 \end{equation}
 and 
 \begin{equation}
  A_{S,R_k} = \frac{4C}{\pi} \int_{0}^{\pi/2} \sin^{4\mu_{S,R_k}}(\theta) {\rm d}\theta - \frac{4C^2}{\pi} \int_{0}^{\pi/4} \sin^{4\mu_{S,R_k}}(\theta) {\rm d}\theta 
 \end{equation}  
respectively. 
Exact closed-form expressions for  $ A_{S,R_k}$ for both modulation schemes can be obtained  by following the same methodology and procedure as in the previous case yielding
\begin{equation}\label{MK3}
A_{S,R_k} =  
\frac{\Gamma(2\mu_{S,R_k} +\frac{1}{2})}{2\sqrt{\pi}\Gamma(2\mu_{S,R_k} +1)}
+\frac{\cos(\pi/M)}{\pi}
 \;_{2}F_{1} \left( \frac{1}{2}-2\mu_{S,R_k} ,\frac{1}{2};  \frac{3}{2}; \cos^2(\pi/M) \right) 
\end{equation}   
for the case of $M-$PSK and  
\begin{equation}
 A_{S,R_k} = 
\frac{2C\Gamma(2\mu_{S,R_k} +\frac{1}{2})}{\sqrt{\pi}\Gamma(2\mu_{S,R_k} +1)}
-
\frac{C^2\Gamma(2\mu_{S,R_k} +\frac{1}{2})\;_{2}F_{1} \left(\frac{1}{2},2\mu_{S,R_k} +\frac{1}{2}; 2\mu_{S,R_k} +\frac{3}{2}; \frac{1}{2}\right)}{\pi2^{2\mu_{S,R_k} +\frac{1}{2}}\; \Gamma(2\mu_{S,R_k} +\frac{3}{2})}
\end{equation}
for the case of $M-$QAM. It is noted here that at sufficiently  high SNR,  the probability $ P(A(k) = C(k) = 1) $ is clearly   smaller than unity and thus, $ 1 - A_{S,R_k}((4\mu_{S,R_k}^{2}h_{S,R_k})/(g^{2} \overline{\gamma}_{S,R_k}^{2}))^{\mu_{S,R_k}} \simeq 1$. 
\newline
To this effect and by substituting  \eqref{LLL45} and \eqref{X} into \eqref{L8}, the    SER of the multi-relay   regenerative system over   generalized fading channels at the  high SNR regime can be  expressed  as follows: 
\begin{equation}\label{LLL51}
P_{\rm SER}^{D}\approx \left(\frac{4\mu_{S,D}^{2}h_{S,D}}{g^{2}\overline{\gamma}_{S,D}^{2}}\right)^{\mu_{S,D}} \sum_{z = 0}^{2^K -1} \prod_{k=1}^{K}A_{R_k,D}(\mathbf{C}_{z}) \left(\frac{4\mu_{R_k,D}^{2}h_{R_k,D}}{g^{2}\overline{\gamma}_{R_k,D}^{2}}\right)^{\mu_{R_k,D}}
\prod_{k=1}^{K} A_{{S,R_k}}
\left(\frac{4\mu_{S,R_k}^{2}h_{S,R_k}}{g^{2}\overline{\gamma}_{S,R_k}^{2}}\right)^{\mu_{S,R_k}} \hspace{0.3cm}
\end{equation}
where $g$ corresponds to  $g_{\rm PSK} = \sin^2(\pi/M)$ or   $ g_{\rm QAM} = 3/(2(M-1))$ according to \eqref{L10} and \eqref{LL32},  respectively, depending on the selected modulation scheme.

\subsection{Amount of Fading }

It is recalled that the amount of fading (AoF) is a  useful metric for quantifying the fading  severity in the  communication scenarios and is defined according to  \cite[eq. (1.27)]{R6}, namely  
\begin{equation}\label{LLLX60}
 \text{AoF} = \frac{\text{Var}(\gamma_{\text{MRC}})}{(\text{E}(\gamma_{\text{MRC}}))^2} = \frac{\text{E}( \gamma_{\text{MRC}}^{2}) - (\text{E}(\gamma_{\text{MRC}}))^{2}}{(\text{E}(\gamma_{\text{MRC}}))^{2}}.
\end{equation} 
The $n^{\rm th}$ moment of the $\gamma_{\text{MRC}}$ under the decode-and-forward strategy can be determined by  \cite{SA}
\begin{equation}\label{LLLLX60}
\mu_{n} = (-1)^{n}\left[\frac{d^{n}}{ds^{n}}\left(  M_{\gamma_{S,D}} (s)  \prod_{k =1}^{K}C(k) M_{R_k,D} (s) \right)\right]_{s = 0}.
\end{equation}
Based on this, the first two moments in \eqref{LLLLX60} are obtained for $n = 1$ and $n = 2$, namely,  $ \text{E}(\gamma_{\text{MRC}}) = 
\partial \text{M}_{\gamma_{\text{MRC}}}(s)/\partial s
|_{s = 0}$ and $\text{E}(\gamma_{\text{MRC}}^{2}) = 
\partial^{2} \text{M}_{\gamma_{\text{MRC}}}(s)/\partial^{2} s
|_{s = 0}.$
To this effect and recalling  \eqref{MGF}${-}$\eqref{L14} as well as carrying out long but basic algebraic manipulations,  the corresponding minimum AoF, if all relays decode correctly,  can be represented as follows:
\begin{equation}\label{LLLX61}
 \text{AoF}  =    \frac{\mu_{S,D}(\mu_{S,D} +1)\delta_{1}^{2}-2\delta_{2}\mu_{S,D} -2\delta_{3} K\mu_{R,D} }{(\mu_{S,D}\delta_{1}+ K\mu_{R,D}\delta_{4})^{2}}  + \frac{    \mu_{R,D}(K\mu_{R,D}+1)\delta_{4}^{2} + 2 \mu_{S,D}\mu_{R,D} \delta_{1}\delta_{4} }{(\mu_{S,D}\delta_{1}+ K\mu_{R,D}\delta_{4})^{2} K^{-1} }      - 1   
\end{equation}
where $\delta_{1} = \overline{\gamma}_{S,D}h_{S,D}/\mu_{S,D}(h_{S,D}^{2} -H_{S,D}^{2}), \delta_{2} = \overline{\gamma}_{S,D}^{2}/4\mu_{S,D}^{2}(h_{S,D}^{2}-H_{S,D}^{2})  
,  \delta_{3} = \overline{\gamma}_{R,D}^{2}/4\mu_{R,D}^{2}(h_{R,D}^{2}\newline -H_{R,D}^{2}) \;\text{and}\; 
\delta_{4} =  \overline{\gamma}_{R,D}h_{R,D}/\mu_{R,D}(h_{R,D}^{2} -H_{R,D}^{2})$.
By recalling that the $\eta-\mu$ model  includes the Nakagami$-m$,  Hoyt and Rayleigh distributions, the   AoF for these cases can be readily deduced.

\section{Optimum Power Allocation}
\indent
This section is devoted to the derivation of the OPA strategy that minimizes the derived asymptotic  SER of the considered regenerative system subject to the sum-power constraint of a power budget $P$.  Since the derived SER expression in \eqref{LLL51} is a function of the power allocated at the source and the $ K $ relays of the system, the available power should be allocated optimally in order to enhance the end-to-end quality of the transmission.  
Based on this, the corresponding non-linear optimization problem can be  formulated as follows: 
\begin{equation}\label{LLL62}
\begin{split}
\mathbf{a_{opt}} = \text{arg} \min_{\mathbf{a}}P_{\rm SER}^{D}\hspace{3.5cm}\\
\text{Subject to}: \quad a_{0} + \sum_{k = 1}^{K}a_{R_k} = 1, \quad a_{0} \geq 0, a_{R_k} \geq 0  \, \qquad  \,  \qquad  \, 
\end{split}
\end{equation}
where $\mathbf{a} = [a_{0},a_{R_1},a_{R_2}, \ldots, a_{R_k}]$  denotes the relative power allocation vector.  Importantly, the cost function in \eqref{LLL62} is convex  in the parameters $a_{0}$ and $a_{R_k}$ over the feasible set defined by linear power ratio constraints. 
The corresponding proof is provided in Appendix II. 
To this effect and following the definitions in \cite{R9},   the Lagrangian of this optimization problem is given by 
\begin{equation}\label{LLL63}
L = P_{\rm SER}^{D} + \lambda\left( a_{0} + \sum_{k=1}^{K} a_{R_k}-1 \right) - \mu_{0}a_{0}-\sum_{k=1}^{K}\mu_{k} a_{R_k}            
\end{equation}
where $\lambda$ denotes the Lagrange multiplier satisfying the power constraint whereas $\mu_0$ and $\mu_k$ parameters serve as slack variables. The nonlinear optimization problem in \eqref{LLL62} can be solved   using e.g. a line search method. 
However, to develop some insights for the power allocation policy we apply the Karush-Kuhn-Tucker  (KKT) conditions for minimizing the corresponding SER \cite{R9}. To this end, it  follows that all  $\mu_k $ and $\mu_0 $ parameters are zero while the following  derivatives   form the necessary condition for maximizing the performance of the system
\begin{equation}\label{LLL64}
\frac{\partial P_{\rm SER}^{D}}{\partial a_{0}} = \frac{\partial P_{\rm SER}^{D}}{\partial a_{R_k}},  \qquad \quad   (1\leq  \: k \leq \: K)
\end{equation}
In order to obtain  feasible relations between optimal powers of the cooperating nodes, we employ  the asymptotic SER in \eqref{LLL51}. By   re-writing this    in terms of the power ratios  allocated at the transmitting nodes   it follows that 
\begin{equation} \label{LLL65}
\begin{split}
P_{\rm SER}^{D}\approx& \left(\frac{4\mu_{S,D}^{2}h_{S,D}N_{0}^{2}}{g^{2}a_{0}^{2}\Omega_{S,D}^{2}P^{2}}\right)^{\mu_{S,D}}  \sum_{z = 0}^{2^K -1} \prod_{k=1}^{K}  A_{R_k,D}(\mathbf{C}_{z}  )\left(\frac{4\mu_{R,D}^{2}h_{R_k,D}N_{0}^{2}}{a_{R_k}^{2}g^{2}\Omega_{R_k,D}^{2}P^{2}}\right)^{\mu_{R_k,D}} 
\\ & \times \prod_{k=1}^{K}A_{S,R_k}\left(\frac{4\mu_{S,R_k}^{2}h_{S,R_k}N_{0}^{2}}{ a_{0}^{2}g^{2}\Omega_{S,R_k}^{2}P^{2}}\right)^{\mu_{S,R_k}}.\hspace{5.5cm}
\end{split}
\end{equation}  
In order to derive an optimal power allocation policy for the DF protocol, we initially restrict our scenario  to $ K = 1 $ and $ K = 2 $ relay nodes and then we generalize the result for  $ K $ relays. 

\subsubsection{K = 1 Scenario}   The possible decoding sets in this case are $ C_{0} = 0 $ i.e., the relay is unable to decode correctly, and $ C_{1}= 1 $ i.e. the relay decodes successfully. Thus, using   \eqref{LLL65} and neglecting  the constant term outside the summation after factoring out $a_{0}$, it follows that  
\begin{equation}\label{L33}
\begin{split}
\min \left[\frac{K_{1}}{a_{0}^{2(\mu_{S,D} + \mu_{S,R_1})}}  + \frac{K_{2}}{a_{0}^{2\mu_{S,D}} a_{R_1}^{2\mu_{R_1,D}}} \right]\\ \vspace{0.5cm} 
\text{Subject to}: a_{0} + a_{R_1} = 1 \hspace{0.8cm}
\end{split}
\end{equation} 
where 
\begin{equation}
K_{1} = A_{S,D}A_{S,R_1}\left(\frac{4\mu_{S,R_1}^{2}h_{S,R_1}N^{2}_{0}}{g^{2}\Omega_{S,R_1}^{2}P^{2}}\right)^{\mu_{S,R_1}} 
\end{equation}
and
\begin{equation}
K_{2} = A_{R_1,D}\left(\frac{4\mu_{R_1,D}^{2}h_{R_1,D}N^{2}_{0}}{g^{2}\Omega_{R_1,D}^{2}P^{2}}\right)^{\mu_{R_1,D}}. 
\end{equation}
Next, we apply the necessary condition in \eqref{LLL64} to determine  the relation between optimal power allocations in the two nodes. Thus, the first  derivative of $ P_{\rm SER}^{D} $ with respect to $a_{0}$ is given by 
\begin{equation}\label{L34}
\frac{\partial P_{\rm SER}^{D}}{\partial a_{0}} = 
-\frac{2(\mu_{S,D} + \mu_{S,R_1}) K_{1}}{a_{0}^{2(\mu_{S,D} + \mu_{S,R_1})+1}} - \frac{2\mu_{S,D}  K_{2}}{a_{0}^{2\mu_{S,D}+1} a_{R_1}^{2\mu_{R_1,D}}} 
\end{equation} 
whereas  the first derivative with respect to $a_{R_1}$ is given by 
\begin{equation}\label{L35}
\frac{\partial P_{\rm SER}^{D}}{\partial a_{R_1}} =   - \frac{2\mu_{R_1,D} K_{2}  }{a_{0}^{2\mu_{S,D}} a_{R_1}^{2\mu_{R_1,D}+1}}. 
\end{equation} 
Equating the above two relations and after some long but basic rearrangements, it follows that
\begin{equation}\label{L36}
\frac{(\mu_{S,D} + \mu_{S,R_1})K_{1}}{K_{2}} = \frac{a_{0}^{2\mu_{S,R_1}+1}}{a_{R_1}^{2\mu_{R_1,D}}}\left( \frac{\mu_{R_1,D}}{a_{R_1}} - \frac{\mu_{S,D}}{a_{0}}\right).
\end{equation}
It is noticed that the left-hand side of \eqref{L36} depends only on the channel parameters.  As a result, this term is  always positive which   implies directly a power policy of $ a_{0}\mu_{R_1,D} \: \geq  \: a_{R_1,D} \mu_{S,D}.$  This relation  is also in agreement with the power allocation  over Rayleigh, Nakagami${-}m$, and Hoyt (Nakagami${-}q$) fading distributions  under the special cases given in section III. 

\subsubsection{$ K = 2 $ Scenario} In this case, the following four scenarios are valid:  $   \mathbf{C}_{0}  = (0,0)$ when the two relays decode incorrectly; $  \mathbf{C}_{1} = (0,1)$ when the first relay decodes incorrectly and the second relay decodes successfully; $\mathbf{C}_{2} = (1,0)$ when the first relay decodes successfully and the second relay decodes incorrectly; and $ \mathbf{C}_{3} = (1,1)$ when both relays decode successfully.  Based on this, the corresponding optimization problem can be expressed as follows

\begin{equation}\label{L38}
\begin{split}
\min \left[\frac{K_{3}}{a_{0}^{2(\mu_{S,D} + \mu_{S,R_1}+\mu_{S,R_2})}}  + \frac{K_{4}}{a_{0}^{2(\mu_{S,D}+\mu_{S,R_1})} a_{R_2}^{2\mu_{R_2,D}}}+\frac{K_{5}}{a_{0}^{2(\mu_{S,D} + \mu_{S,R_2})}a_{R_1}^{2\mu_{R_1,D}}}  + \frac{K_{6}}{a_{0}^{2\mu_{S,D}} a_{R_2}^{2\mu_{R_2,D}} a_{R_1}^{2\mu_{R_1,D}}} \right]  \\
\end{split}
\end{equation}
\begin{equation*}
\hspace{6cm} \text{Subject to}: a_{0} + a_{R_1} + a_{R_2} = 1 \hspace{6cm}
\end{equation*}
where $ K_{3},K_{4}, K_{5} \: \text{and} \: K_{6}$ relate to the channel parameters, which are not affecting the sign of the derivatives in any case.  To this effect, the derivative of $ P_{\rm SER}^{D} $ with respect to $ a_{R_1} $ is given by  
\begin{equation}\label{L39}
\frac{\partial P_{\rm SER}^{D}}{\partial a_{R_1}} = - \frac{2K_{5}\mu_{R_1,D}}{a_{0}^{2(\mu_{S,D} + \mu_{S,R_2})}a_{R_1}^{2\mu_{R_1,D}+1}} - \frac{2K_{6}\mu_{R_1,D}}{a_{0}^{2\mu_{S,D}} a_{R_2}^{2\mu_{R_2,D}} a_{R_1}^{2\mu_{R_1,D}+1}}
\end{equation}
whereas the derivative of  $ P_{\rm SER}^{D} $  with respect to $ a_{R_2} $ is given by 
\begin{equation}\label{L40}
\frac{\partial P_{\rm SER}^{D}}{\partial a_{R_2}} =  - \frac{2K_{4}\mu_{R_2,D}}{a_{0}^{2(\mu_{S,D} + \mu_{S,R_1})}a_{R_2}^{2\mu_{R_2,D}+1}} - \frac{2K_{6}\mu_{R_2,D}}{a_{0}^{2\mu_{S,D}} a_{R_2}^{2\mu_{R_2,D}+1} a_{R_1}^{2\mu_{R_1,D}}}. 
\end{equation}
With the aid of the  above two representations, a power allocation strategy can be proposed for identical channel conditions as $ a_{R_1} = a_{R_2}.$ Likewise, applying $ \partial P_{\rm SER}^{D}/\partial a_{R_k} =  \partial P_{\rm SER}^{D}/\partial a_{R_k+1}$, $(2\leq k \leq K) $ it is straightforwardly  shown that $ a_{R_k} =  a_{R_k+1} $. This power assignment indicates that under the total power constraint and the assumed channel conditions for the regenerative network, power control mechanism with the relays is not essential, but the remaining power left from the total power budget after the source can be simply allocated uniformly between the nodes. This is further elaborated and largely verified in the following section.  

\begin{figure}[bp!]
\begin{minipage}{0.45\linewidth}
\includegraphics[width=83mm, height = 83mm]{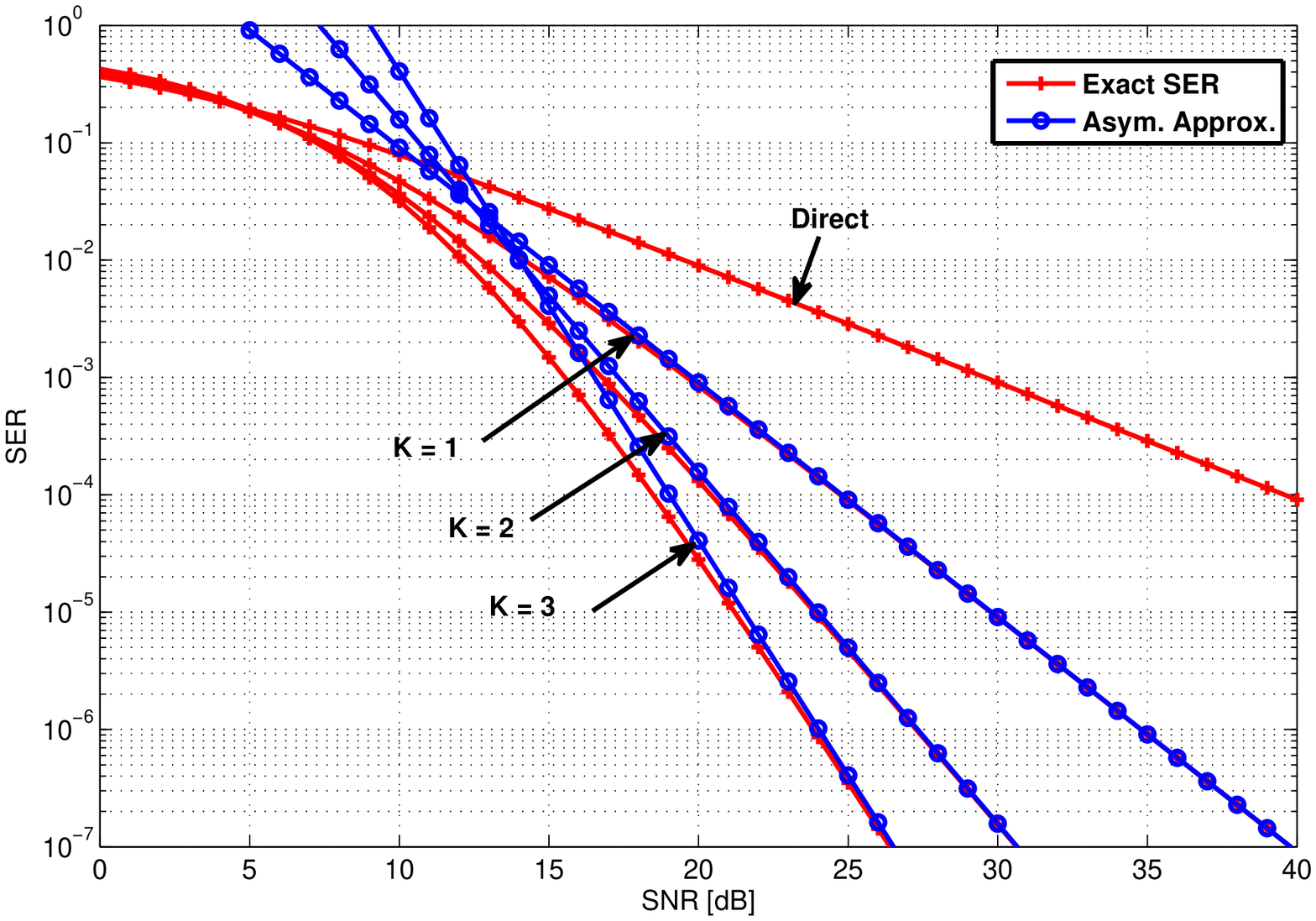} 
\caption{SER  in  $\eta-\mu $ fading  with $ \mu = 0.5, \eta = 1$ and   $\Omega_{S,D} = \Omega_{S,R_k} = \Omega_{R_k,D} = 0$dB \;for 4${-}$PSK/$4-$QAM constellation with different number of relays and EPA.} 
\end{minipage}
\hfill
\begin{minipage}{0.45\linewidth}
\centering
\includegraphics[width=83mm, height=83mm]{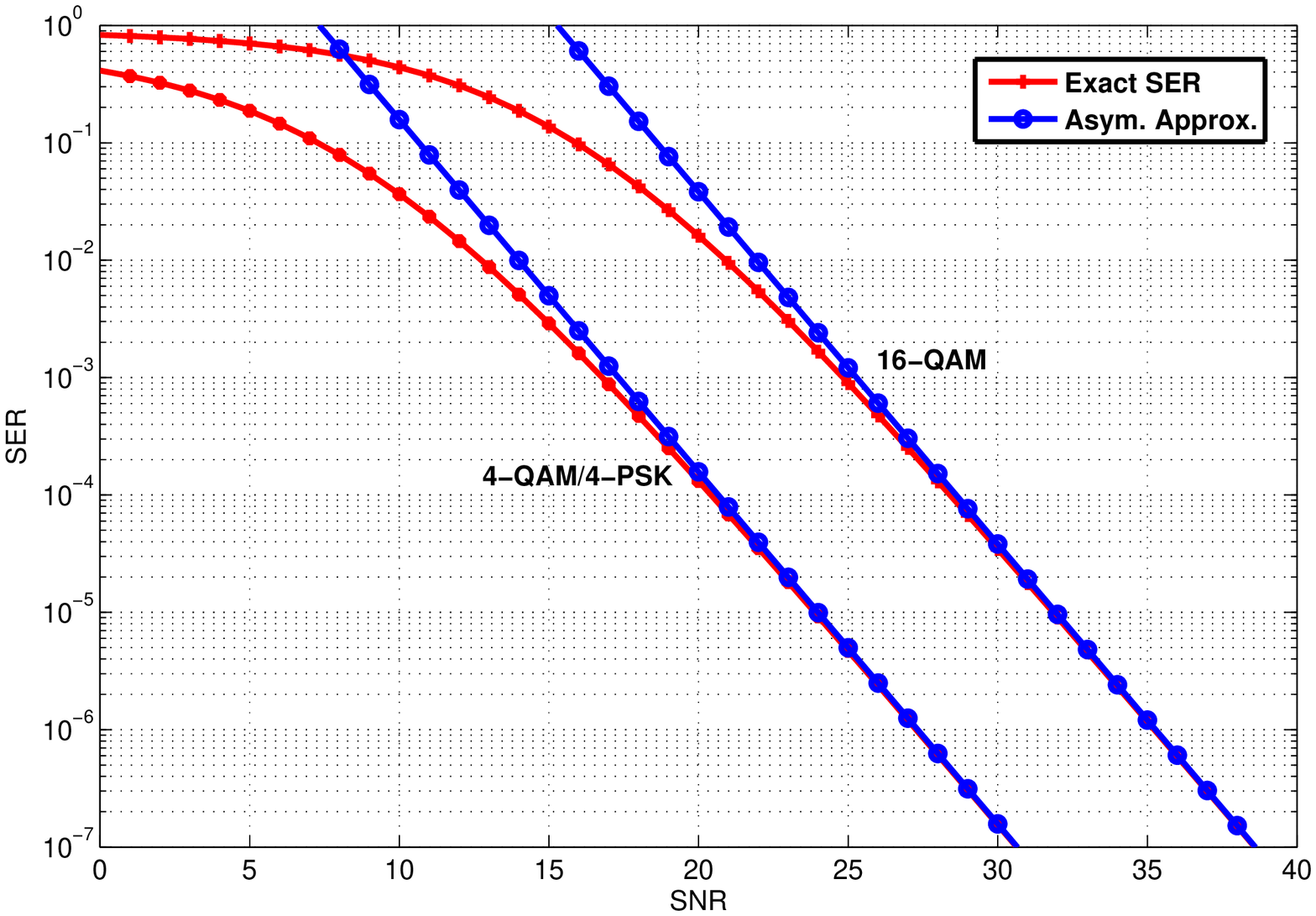}
\caption{SER  in  $\eta -\mu $ fading   with  $ \mu = 0.5, \eta = 1 \;\text{and}\; \Omega_{S,D} = \Omega_{S,R_k} = \Omega_{R_k,D} = 0$dB \;for $4-$QAM/$4-$PSK and  16${-}$QAM, $K = 2$ and EPA.}
\end{minipage} 
\end{figure}
   
\section{Numerical Results}

In this Section, the offered analytic results are employed in evaluating the performance of the considered regenerative system for different communication scenarios. 
To this end, the variance of the noise is assumed to be normalized as $ N_{0} = 1 $ for all considered scenarios while transmit powers are equally or optimally allocated to the source and the relays. It is also noted that  the presented results are limited to Format$-1$ of the $\eta-\mu$ distribution but they can be   readily extended to scenarios that correspond to  Format$-2$  \cite{R1}.

\begin{figure}[bp!]
\begin{minipage}{0.45\linewidth}
\includegraphics[width=83mm, height = 83mm]{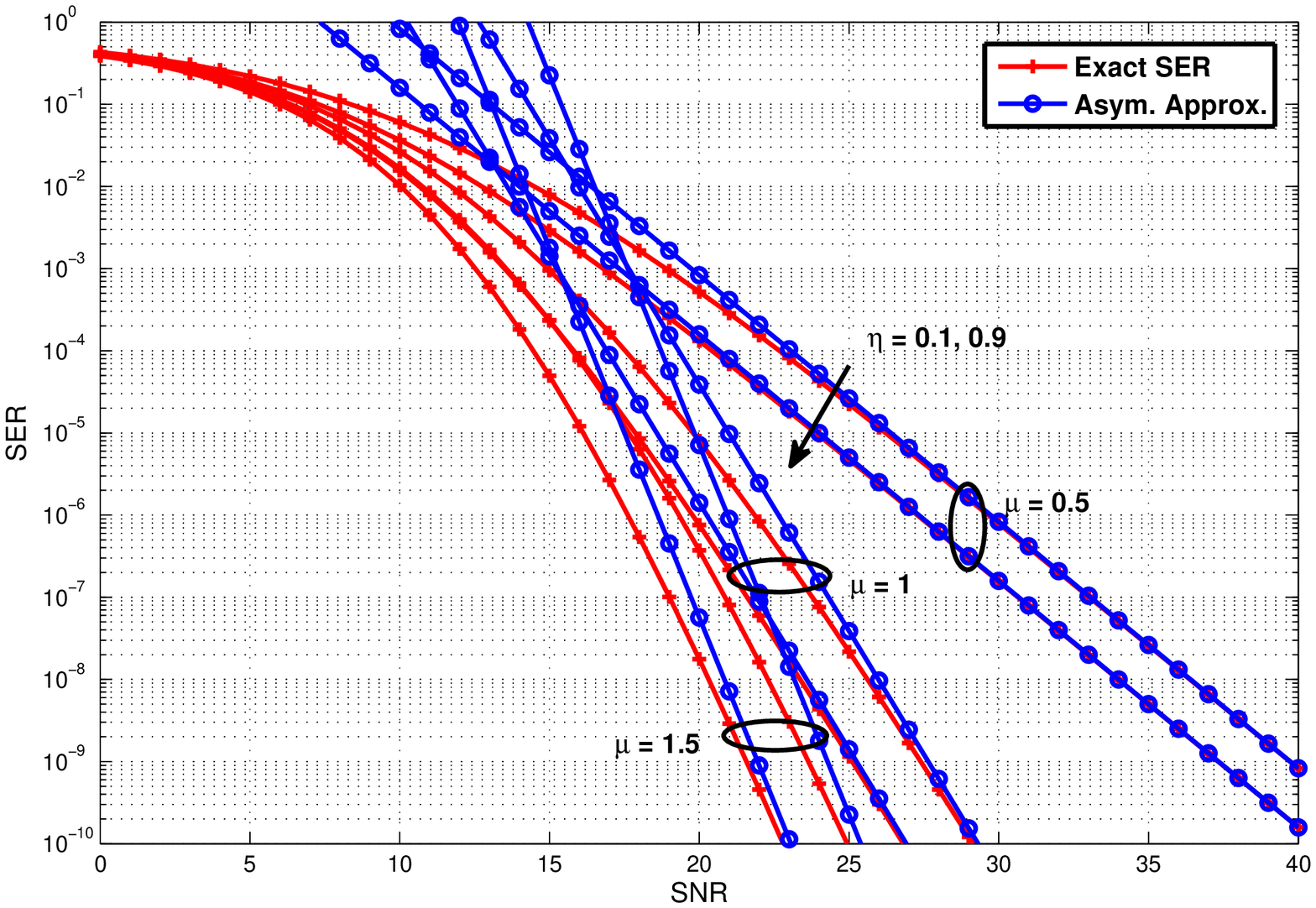} 
\caption{SER  vs average SNR with $ \mu = \{ 0.5, 1, 1.5 \}$, $\eta = \{ 0.1, 0.9\}$ and $ \Omega_{S,D} = \Omega_{S,R_k} = \Omega_{R_k,D} = 0$dB for $4-$PSK/4${-}$QAM,  $K=2$ and EPA.} 
\end{minipage}
\hfill
\begin{minipage}{0.45\linewidth}
\centering
\includegraphics[width=83mm, height=83mm]{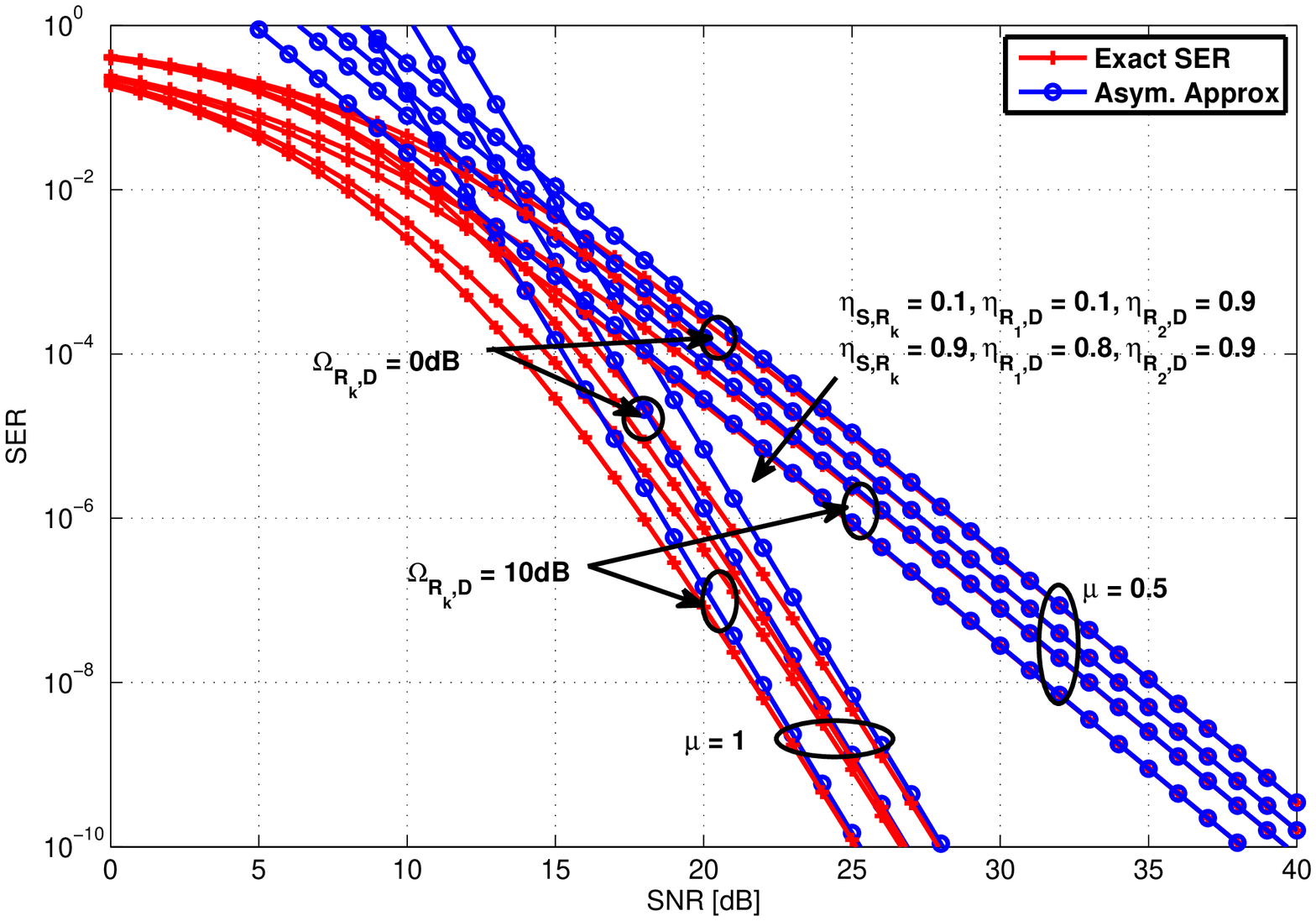}
\caption{SER   vs average SNR with $ \mu = \{ 0.5, 1  \}$, $\eta_{S,D} = 0.9$, $\Omega_{S,D} = \Omega_{S,R_k} =  0$dB, $ \Omega_{R_k,D} = \{ 0, 10 \} $dB with diffferent $\eta_{S,R_k} \;\text{and}\; \eta_{R_1,D} $ for $4-$PSK/4${-}$QAM,  $ K = 2 $ and EPA.}
\end{minipage} 
\end{figure}

Fig. $2$ illustrates the SER performance as a function of SNR for  one, two and three  relays using equal power allocation, i.e., $ P_{0} = P_{R_k} = P/( K+1)$  over symmetric and balanced $\eta-\mu$ fading channels for  4${-}$PSK$/$4$-$QAM constellation. 
Also,  the $\eta-\mu$ fading parameters are  $\mu  = 0.5$ and $\eta $ = 1 while $\Omega$ parameters are equal to unity. 
It is shown that the exact results are bounded tightly  by the corresponding asymptotic curves from moderate to high SNR values while a  full diversity order i.e., 2, 3 and 4, can be achieved. 
Table II depicts the corresponding diversity gains computed from the slopes for the exact and asymptotic curves along with the direct transmission scenario, for reference,  where it is assumed that  $P_{0} = P $.  It is observed that  at a target SER of $10^{-4}$ the single relay   system exhibits a gain of $15$dB   over the direct transmission  whereas the two and three relay  systems outperform the direct  scenario by about $19.5$dB and $21.5$dB, respectively.   
In the same context, Fig. 3 illustrates  the exact and asymptotic results for $4{-}$QAM/$4-$PSK and $16-$QAM constellations  for the case of two relays  with equal power allocation over symmetric  $\eta-\mu $ fading channels with $\mu = 0.5 $ and $\eta = 1 $ as well as  balanced links i.e., $\Omega_{S,D} = \Omega_{S,R_k} = \Omega_{R_k,D} = \Omega = 0$dB.   
 It is shown that the asymptotic   curves are almost identical to the exact  ones for SERs lower than around $10^{-3}$.  Therefore, it becomes evident that in practical system designs of  DF relaying at the  high SNR regime, the offered asymptotic expressions can provide useful insights on the system performance. 
\begin{table}[tp!]
\centering
\caption{diversity gains for one, two and three relays using 4${-}$PSK/4$-$QAM for $\mu$ = 0.5 and $\eta$ = 1.}
\begin{tabular}{|c|c|c|}
\hline\hline
$ K $ & Diversity Gain (Exact) & Diversity Gain (Asymp.) \\
\hline
1 & 1.96 &2\\
\hline
2  & 2.91 & 3   \\
\hline
3 & 3.85 & 4   \\
\hline
\hline
\end{tabular}
\end{table}
 %
\begin{figure}[tbp!]
\centering
\includegraphics[width=120mm, height=70mm]{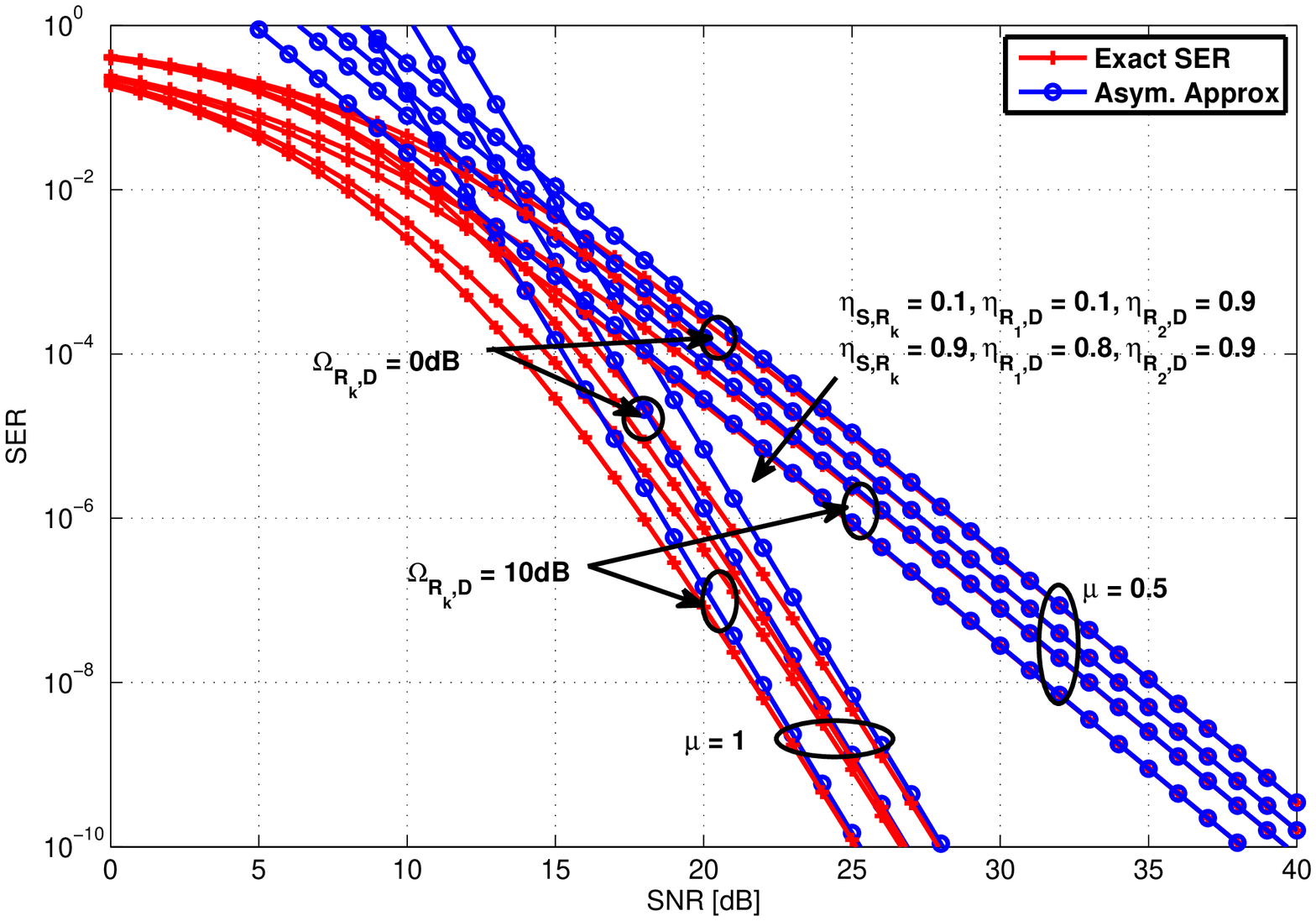}
\caption{SER   in  $\eta -\mu $ fading with $ \mu = \{ 0.5, 1  \}$, $\eta_{S,D} = 0.9$ and $\Omega_{S,D} = \Omega_{S,R_k} =  0$dB, $ \Omega_{R_k,D} = \{ 0, 10 \} $dB with diffferent $\eta_{S,R_k} \;\text{and}\; \eta_{R_1,D} $ for $4-$PSK/4${-}$QAM signals,  $ K = 2 $ and EPA.}
\end{figure}

Fig. 4 illustrates  the cooperation performance of 4${-}$QAM/QPSK system  in a two relay scenario for $\mu = \{ 0.5, 1, 1.5\}$ and identical channel variance of $ \Omega_{S,D} = \Omega_{S,R_k} = \Omega_{R_k,D} = 0$dB with equal power allocation. It is also recalled that the case of  $\mu = 0.5 $ corresponds to the Nakagami${-}q$  (Hoyt)\;distribution with $q^{2} = \eta $. 
By varying the value of $\eta$,  we observe the effect of   the  scattered-wave  power ratio on the average SER of the considered regenerative system. 
This verifies that  the SER is inversely proportional to $\eta$ since for an indicative  SER of $10^{-4}$, an average gain of 2dB is observed when $\eta$ increases from 0.1 to 0.9 for all values of $ \mu $.

Furthermore,  average gains of 4dB and 1.75dB are obtained as   $ \mu $ increases from 0.5 to 1 and from 1 to 1.5, respectively. 
Likewise, Fig. 5, demonstrates the SER performance in i.n.i.d  $\eta-\mu$ fading channels  for 4${-}$QAM/4$-$PSK constellations  for the case of two relays with equal power allocation.
It is assumed that $\eta_{S,D} = 0.9$, $\Omega_{S,D} = \Omega_{S,R_k} =  0$dB, $ \Omega_{R_k,D} = \{0, 10\}$dB, $\mu_{S,D} = \mu_{S,R_k} = \mu_{R_k,D} = \mu  = \{0.5,1\}$ with different values of $\eta_{S,R_k} $ and $\eta_{R_1,D}$  set as $\{ \eta_{S,R_k} , \eta_{R_1,D}, \eta_{R_2,D} \} = \{0.1, 0.1, 0.9 \}$ and $ \{0.9, 0.8, 0.9 \}$, respectively. It is observed that the performance of the system improves substantially as  $\eta_{S,R_k} $ and   $\eta_{R_1,D} $ increase as at a SER of $10^{-4}$  almost 1.25dB and 1.75dB gains are achieved when $\{ \eta_{S,R_k}, \eta_{R_1,D}, \eta_{R_2,D} \}$ changes from $ \{0.1, 0.1, 0.9 \}$ to $ \{0.9, 0.8, 0.9 \}$ for $\Omega_{R_k,D} = \{0, 10 \} $dB, for the considered values of $\mu$.    
 %
\begin{figure}[tbp!]
\centering
\includegraphics[width=120mm,height=70mm]{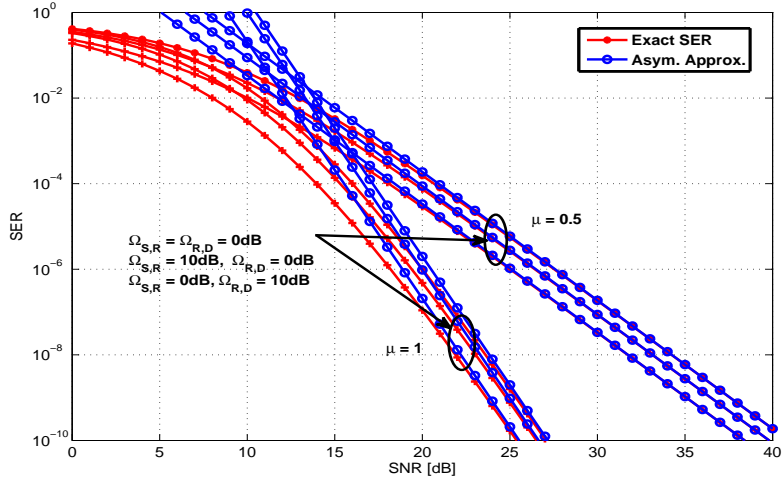}
\caption{SER in  $\eta-\mu$ fading  with $\mu = \{ 0.5, 1 \}$, $\eta = 0.5$, and $\Omega_{S,D} = 0$dB for different $\Omega_{S,R_k}$ and $\Omega_{R_k,D}$,  4${-}$QAM/$4-$PSK,  $K = 2$ and EPA.} 
\end{figure}
 
 Fig. 6 illustrates  the SER performance in $ \eta-\mu $ fading conditions with $ \mu_{S,D} = 0.5 \; \text{and}\; \eta = 0.1 $  for  4${-}$QAM/$4-$PSK constellations and balanced links of relative channel variance   0dB for the case of two relays using equal power allocation with different $\mu_{S,R_k}$ and non-identical values of $ \mu_{R_k,D}$. The figure shows that increasing at least one of $ \mu_{R_k,D}$'s value at a fixed $\mu_{S,R_k}$ or increasing  both  $\mu_{S,R_k}$  and   $ \mu_{R_k,D} $ simultaneously can improve the cooperation performance. Indicatively, at SER of $10^{-4} $ nearly 1.25dB and 1.75dB gains are observed when $\{\mu_{S,R_k},\mu_{R_1}, \mu_{R_2}    \}$ changes from $ \{0.5, 0.5, 0.5 \}  $ to $ \{0.5, 0.5, 1 \}  $ and  from  $ \{0.5, 0.5, 1 \}  $ to $ \{0.5, 1, 1.5 \}  $, respectively.  Also, nearly 0.75dB and 1dB gains are achieved when $ \{\mu_{R_1}, \mu_{R_2}\}  $ varies from $ \{1, 1 \}  $ to $ \{1, 1.5 \}$ and then to $ \{1.5, 2 \}$ when $ \mu_{S, R_k} = 1 $, whereas a gain of 4dB is shown when both  $\mu_{S,R_k}$ and $\mu_{R_k, D}$ increase at the same time, for instance, from $ \{0.5, 0.5, 1\} $ to $ \{1, 1, 1.5\} $. Likewise, Fig. 7,  depicts the SER performance for 4${-}$QAM/$4-$PSK constellations for  $\mu = 0.5\;\text{and}\;1$, $\eta = 0.5$ and unbalanced channel links employing two relays with equal power allocation. It is shown that  increasing either $\Omega_{S,R_k}$ or $\Omega_{R_k,D}$ can improve the SER  and  that the performance for the case of   $\Omega_{S,R_k} = 0$dB and  $\Omega_{R_k,D} = 10$dB is better than the reverse scenario i.e.  $ \Omega_{S,R_k} = 10$dB and  $\Omega_{R_k,D} = 0$dB.  
  For example,  almost constant gains of 1dB and 1.75dB are achieved when $ \Omega_{S,R_k} = 0$dB and $\Omega_{R_k,D} = 0$dB increase to $ \Omega_{S,R_k} =$10dB and $\Omega_{R_k,D} = 0$dB and from $ \Omega_{S,R_k} =$0dB and $\Omega_{R_k,D} = 0$dB to $ \Omega_{S,R_k} =$0dB and $ \Omega_{R_k,D} = 10$dB, respectively.   This verifies that  in the considered regenerative protocol, the overall performance improves more  when increasing  average power from the relays to the destination than from the source to the relays. 
\begin{figure}[tbp!]
\centering
\includegraphics[width=120mm, height=70mm]{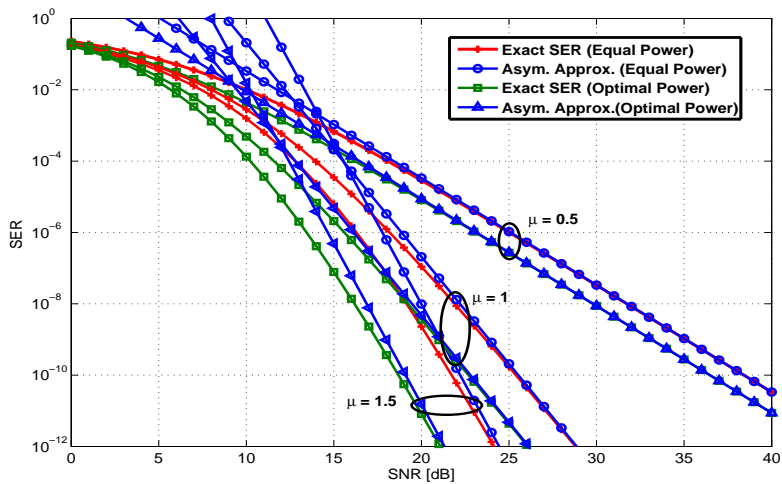}
\caption{SER with OPA in  $\eta-\mu $ fading with  $ \eta = 0.5$, $\Omega_{S,D} = \Omega_{S,R_k} =0$dB and $\Omega_{R_k,D} =10$dB for different $ \mu$  and 4${-}$QAM/$4-$PSK  and $K=2$.}
\end{figure}
\hfill

\begin{table}[htbp!]
\caption{optimal transmit power allocations with different \:$ \mu_{S,R_k} $ \:and \: $ \mu_{R_k,D} $ \:for\: $\mu_{S,D} = 0.5 $, $  \eta = 0.5, \Omega_{S,D} = \Omega_{S,R_k}  = \Omega_{R_k,D}  = 0 ${\rm d}B  \:using $4{-}$QAM/$4-$PSK \:signals at SNR = 20 {\rm d}B.}  
\centering
\centering
\begin{tiny}
\begin{tabular}{|c|c|c|c|}
\hline
$ \mu_{S,R_k} $ & $ \mu_{R_k,D} $  & $ P_{0}/P \,\, | \,\, P_{R_1}/P $ & $ P_{0}/P \,\, | \,\, P_{R_1,R_2}/P $  
\\
\hline\hline     
$  $ & $ 0.5  $  & $ 0.6270     \,\, | \,\, 0.3730 $ & $ 0.4832     \,\, | \,\,  0.2584 $  \\
\cline{2-4}
$ 0.5 $ & $ 1  $  & $ 0.6572  \,\, | \,\,0.3428  $ & $ 0.4914\,\, | \,\,  0.2543 $ \\
\cline{2-4}
$  $ & $ 1.5  $  & $ 0.6871 \,\, | \,\, 0.3129 $ & $ 0.5048 \,\, | \,\,  0.2476 $ \\
\hline
$  $ & $ 0.5  $  & $ 0.5415 \,\, | \,\, 0.4585 $ & $ 0.4006 \,\, | \,\,  0.2997 $  \\
\cline{2-4}
$ 1 $ & $ 1  $  & $ 0.5302 \,\, | \,\, 0.4698$ & $ 0.3842\,\, | \,\,  0.3079 $  \\
\cline{2-4}
$  $ & $ 1.5  $  & $ 0.5725 \,\, | \,\, 0.4275 $ & $ 0.3971 \,\, | \,\,  0.3014 $  \\
\hline
$  $ & $ 0.5  $  & $ 0.5160 \,\, | \,\, 0.4840 $ & $ 0.3724\,\, | \,\,  0.3138 $  \\
\cline{2-4}
$  1.5 $ & $ 1  $  & $ 0.4652\,\, | \,\, 0.5348 $ & $ 0.3443 \,\, | \,\,  0.3278 $ \\
\cline{2-4}
$  $ & $ 1.5  $  & $ 0.5008 \,\, | \,\, 0.4992 $ & $ 0.3424 \,\, | \,\,  0.3288 $  \\
\hline
\hline
\end{tabular}
\end{tiny}
\label{table:System}
\end{table} 

Fig. 8, demonstrates the SER performance of the proposed optimal power allocation scheme for  different values of the fading parameter $\mu$,  and assuming stronger channel variance from the relay nodes to the destination node with constant scattered-wave power ratio  for the case of two relays and 4${-}$QAM/$4-$PSK constellations. It is also assumed that $ \mu_{S,D} = \mu_{S,R_k} = \mu_{R_k,D} = \mu $,\;$ \Omega_{S,D} = \Omega_{S,R_k} =0 $dB, $ \Omega_{R_k,D} = 10$dB and $ \eta = 0.5 $.  
  It is shown that when $ \mu $ is small, for example  $ \mu = 0.5 $, the  OPA provides small gain to the cooperation system, which, however,  increases as $\mu$ increases.  
   Indicatively, for a SER of $ 10^{-4}$  the optimal system outperforms the equal power allocation scenario by at least 1.5dB when $\mu = 0.5 $ and by   2.5dB and 3dB when $ \mu = 1 \;\text{and}\; \mu = 1.5 $, respectively.  
   In addition,  Fig. 9 depicts the corresponding SER for the case that  $ \mu_{S,D} = \mu_{S,R_k} = \mu_{R_k,D} = \mu  $, $ \eta = 0.5 $, $ \Omega_{S,D} = \Omega_{S,R_k} =0 $dB and $ \Omega_{R_k,D} = \{ 0, 10 \}$dB. 
   It is observed that when  $\Omega_{S,R_k} = \Omega_{R_k,D}  = 0 $dB,  OPA does not provide significant performance improvement for the considered  DF network. On the contrary, the SER improves as the  difference between  $\Omega_{S,R_k} $ and $ \Omega_{R_k,D} $  increases.  For example, it is noticed that for a SER of  $10^{-4}$ and $\mu = 0.5$, gains of 0.5dB and 2dB are achieved by OPA over the EPA scheme when  $ \Omega_{R_k,D} = \{ 0, 10 \}$dB,  respectively.  Similarly, gains of 0.5dB and 3dBs are obtained when $\mu$ is increased to $\mu = 1.5$ while it is generally noticed that  OPA is typically more effective than EPA, even in the low-SNR regime.

\begin{figure}[tbp!]
\centering
\includegraphics[width=120mm, height = 70mm]{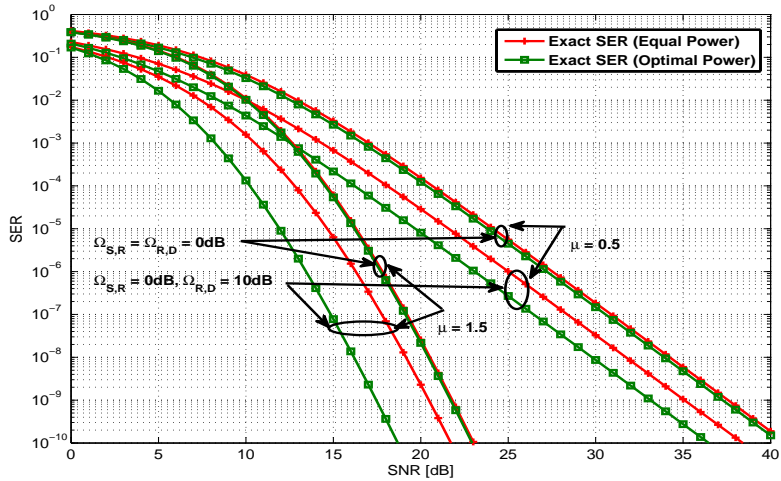} 
\caption{SER with OPA  over $\eta-\mu $ fading  for $ \mu_{S,D} = \mu_{S,R_k} = \mu_{R_k,D} = \mu, \eta = 0.5, \Omega_{S,D} = \Omega_{S,R_k} = 0$dB with different $\Omega_{R_k,D}$  for 4${-}$QAM/$4-$PSK and $K=2$.}
\end{figure}

 \begin{table}[htbp!]
\caption{optimal transmit power allocations with different  $ \Omega_{R_k,D} \:$ \text{and}\:   $\mu_{S,D} = \mu_{S,R_k} = \mu_{R_k,D} = \mu $\:  
\:$ \text{for} \,  \eta = 0.5, \Omega_{S,D} = \Omega_{S,R_k}  = 0 ${\rm d}B  and  $4{-}$ QAM/$4-$PSK \:signals at SNR = 20 {\rm d}B.} 
\centering
\begin{tiny}
\begin{tabular}{|c|c|c|c|c|}
\hline
$ \mu $ & $ \Omega_{R_k,D} $  & $ P_{0}/P \,\, | \,\, P_{R_1}/P $ & $ P_{0}/P \,\, | \,\, P_{R_1,R_2}/P $ &   $ P_{0}/P \,\, | \,\, P_{R_1,R_2,R_3}/P $  
\\
\hline\hline     
$  $ & $ 1  $  & $ 0.6270    \,\, | \,\, 0.3730 $ & $ 0.4832    \,\, | \,\,0.2584 $ & $ 0.4036   \,\, | \,\, 0.1988  $ \\
\cline{2-5}
$ 0.5 $ & $ 10  $  & $ 0.7968     \,\, | \,\, 0.2032$ & $ 0.6974     \,\, | \,\,  0.1513 $ & $ 0.6328    \,\, | \,\,0.1224  $ \\
\cline{2-5}
$  $ & $ 100  $  & $ 0.9181     \,\, | \,\, 0.0819 $ & $ 0.8712    \,\, | \,\,   0.0644$ & $ 0.8371     \,\, | \,\, 0.0543 $ \\
\hline
$  $ & $ 1  $  & $ 0.5925 \,\, | \,\, 0.4075 $ & $ 0.4368    \,\, | \,\,  0.2816 $ & $  0.3520   \,\, | \,\, 0.2160 $ \\
\cline{2-5}
$ 1 $ & $ 10  $  & $0.8316     \,\, | \,\,0.1684 $ & $ 0.7343    \,\, | \,\,  0.1328 $ & $ 0.6658     \,\, | \,\, 0.1114 $ \\
\cline{2-5}
$  $ & $ 100  $  & $ 0.9557     \,\, | \,\, 0.0443 $ & $ 0.9247     \,\, | \,\, 0.0376  $ & $  0.8995     \,\, | \,\, 0.0335 $ \\
\hline
$  $ & $ 1  $  & $  0.5735  \,\, | \,\,  0.4265 $ & $  0.4131     \,\, | \,\, 0.2935  $ & $  0.3274     \,\, | \,\, 0.2242 $ \\
\cline{2-5}
$  1.5 $ & $ 10  $  & $ 0.8496 \,\, | \,\, 0.1504 $ & $ 0.7549     \,\, | \,\,  0.1226 $ & $  0.6850   \,\, | \,\, 0.1050 $ \\
\cline{2-5}
$  $ & $ 100  $  & $ 0.9683    \,\, | \,\, 0.0317 $ & $ 0.9439 \,\, | \,\,  0.0280 $ & $ 0.9232     \,\, | \,\, 0.0256 $ \\
\hline
\hline
\end{tabular}
\end{tiny}
\end{table}

Fig. 10 demonstrates the SER performance of the derived OPA and EPA scenarios for single, two and three relays    over symmetric $\eta-\mu $ fading scenario, i.e., with  constant $\mu $ and  constant scattered-power ratios $\eta$ and unbalanced channel variances from source-to-relay and from relay-to-destination. 
It is assumed that  $\mu_{S,D} = \mu_{S,R_k} = \mu_{R_k,D} = 1 , \eta = 0.5 $ whereas $ \Omega_{S,D} = \Omega_{S, R_k} = 0$dB and $ \Omega_{R_k,D} = 10$dB. 
It is shown that the OPA strategy clearly outperforms its EPA counterpart since the gain for a SER of $ 10^{-4}$ is 2dB, 2.5dB and 2.75dB for one, two and three relays, respectively.   
 The characteristics of the OPA strategy are further analyzed with the aid of  Tables III, IV and V, which depict  the optimal power ratios allocated to the source and the relay-nodes in terms of $ P_{0}/P$ and $ P_{R_k}/P $.  In case of multiple relays, the relays are assigned with equal powers  ($ P_{R_k}/P = P_{R_k+1}/P $).  The power ratios allocated to the source and relay nodes for asymmetric and balanced channel conditions are   tabulated indicating that the numerical values are in tight agreement with the formulated power strategy   in section IV. Table IV also corresponds to the case that the source is assigned with high proportion of power when $\Omega _{R_k,D}$ is larger than $\Omega _{S, R_k}$,  i.e., unbalanced channel links whereas Table V verifies that the optimal power allocation scheme is  dependent upon the considered modulation  scheme. 
\begin{table}[htbp!]
 \caption{optimal transmit power allocations with different modulations and  \: $ \mu_{S,D} = \mu_{S,R_k} = \mu_{R_k,D} = \mu , \eta = 0.5, \Omega_{S,D} = \Omega_{S,R_k}  = \Omega_{R_k,D}  = 0 ${\rm d}B for  two relays at SNR = 20 {\rm d}B.} 
\centering
\begin{tiny}
\begin{tabular}{|c|c|c|c|}
 \hline 
  & 4${-}PSK$ & 16${-}PSK$ & 16${-}QAM$ \\ 
 \hline 
  $\mu $ & $ P_{0}/P \,| \,P_{R_1,R_2}/P $ & $ P_{0}/P \,|\, P_{R_1,R_2}/P $ & $ P_{0}/P \, |\, P_{R_1,R_2}/P $ \\ 
 \hline \hline 
 0.5 & $0.4832\,|\,0.2584$ & $0.4932\,|\, 0.2534$ & $0.5113\,|\, 0.2443$ \\ 
 \hline 
 1 & $ 0.4368\,|\, 0.2816$ & $0.4392\,|\, 0.2804$ & $0.4572\,|\, 0.2714 $ \\ 
 \hline 
 1.5 & $ 0.4130\,|\, 0.2935$ & $ 0.4138\,|\, 0.2931 $ & $ 0.4287\,|\, 0.2857 $ \\ 
 \hline \hline 
 \end{tabular}  
\end{tiny}
\end{table}
\begin{figure}[tbp!]
\centering
\includegraphics[width=120mm, height = 70mm]{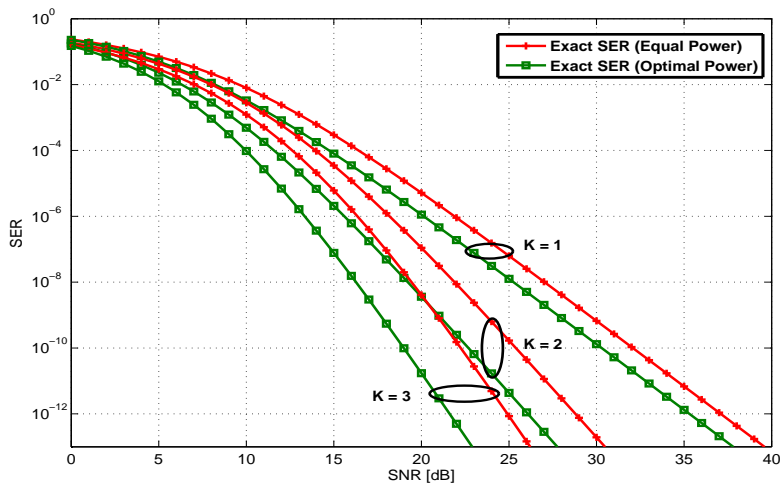} 
\caption{SER with OPA over $\eta-\mu$ fading for $\mu = 1$, $\eta = 0.5$, $\Omega_{S,D} = \Omega_{S,R_k} =0$dB and  $\Omega_{R_k,D} = 10$dB for $4-$PSK/4${-}$QAM and different number of relays.} 
\end{figure}

 \vspace{-0.3333cm}

\section{Conclusion}

In this paper, we   analyzed the  end-to-end performance and optimum power allocation  of   regenerative cooperative systems over   generalized fading channels.     Novel exact and asymptotic closed-form expressions for the  SER assuming $M{-}$PSK and $M{-}$QAM modulated signals were derived over  independent and identically distributed as well as independent and non-identically distributed channels.  
The derived analytic expressions were then used  to draw insight of the different fading parameters in the generalized $\eta-\mu$ fading conditions and their impact on the end-to-end system performance.
 The offered results were subsequently employed in developing an optimum power allocation scheme   which was shown to significantly outperform conventional equal power allocation strategy.  It was also shown that the optimum power allocation scheme is practically independent of  the   scattered-waves power ratio parameter from source-to-destination, while it is dependent upon the number of  multipath clusters as well as  the selected modulation scheme and that it overall  provides significant performance enhancement.  
 
 \vspace{-0.05cm}

\section*{Appendix I \\ A MATLAB Algorithm for Computing the Generalized Lauricella Function}

The Generalized Lauricella function is defined by the following non-infinite single integral, 
\begin{equation}
F_{D}^{(n)}(\mathbf{\rm a}, b_{1}, \cdots, b_{n}, c; x_{1}, \cdots, x_{n}) \triangleq  \frac{\Gamma(c)}{\Gamma(\mathbf{\rm a}) \Gamma(c - \mathbf{\rm a})} \int_{0}^{1} t^{\mathbf{\rm a} - 1} \frac{(1 - t)^{c - \mathbf{\rm a} - 1}}{ (1 - x_{1}t)^{b_{1}} \cdots (1 - x_{n}t)^{b_{n}}} {\rm d}t 
\end{equation}
The numerical evaluation of the above representation was also discussed in \cite[Appendix V]{Jose} and can be straightforwardly evaluated  with the aid of the  following proposed MATLAB algorithm:  
\begin{verbatim}
  Function FD = Lauricella(a, b1, ... , bn, c, x1, ... , xn);  
  f = gamma(c).*gamma(a).*gamma(c - a); 
  Q = @(t)  f.*t.^(a - 1).*(1 - t).^(c - a - 1).*...  
  (1 - x1.*t).^(-b1) ... (1 - xn.*t).^(-bn); 
  FD = quad(Q,0,1)
\end{verbatim}

\vspace{-0.5cm}

 \section*{Appendix II \\ Proof of Convexity of the Optimization Problem}

We provide the proof for the convexity of the SER expression by using \eqref{LLL51}. For mathematical tractability, we consider the proof for three relay-nodes. Based on this,   the proof for larger number of nodes scenario follows immediately. To this end, the asymptotic SER can be expressed as  
\begin{equation*} 
\begin{split}
P_{\rm SER}^{D}\approx &  \frac{K_{1}}{a_{0}^{2(\mu_{S,D} + \mu_{S,R_1}+\mu_{S,R_2}+\mu_{S,R_3})}} + \frac{K_{2}}{a_{0}^{2(\mu_{S,D} + \mu_{S,R_1}+ \mu_{S,R_2})}a_{R_3}^{2\mu_{R_3,D}}}   +   \frac{K_{3}}{a_{0}^{2(\mu_{S,D} +  \mu_{S,R_1}+\mu_{S,R_3})}a_{R_2}^{2\mu_{R_2,D}}} \\
&+  \frac{K_{4}}{a_{0}^{2(\mu_{S,D} + \mu_{S,R_1})}a_{R_2}^{2\mu_{R_2,D}}a_{R_3}^{2\mu_{R_3,D}}} +\frac{K_{5}}{a_{0}^{2(\mu_{S,D} + \mu_{S,R_2}+\mu_{S,R_3})}a_{R_1}^{2\mu_{R_1,D}}} + \frac{K_{6}}{a_{0}^{2(\mu_{S,D}  +\mu_{S,R_2})}a_{R_1}^{2\mu_{R_1,D}}a_{R_3}^{2\mu_{R_3,D}}} 
\end{split}
\end{equation*}
\begin{equation}\label{AB} 
+   \frac{K_{7}}{a_{0}^{2(\mu_{S,D} + \mu_{S,R_3})}a_{R_1}^{2\mu_{R_1,D}}a_{R_2}^{2\mu_{R_2,D}}} + \frac{K_{8}}{a_{0}^{2\mu_{S,D}}a_{R_1}^{2\mu_{R_1,D}}a_{R_2}^{2\mu_{R_2,D}}a_{R_3}^{2\mu_{R_3,D}}}
\end{equation} 
 where $K_{1},\cdots,K_{8} $ are related to the channel parameters. Let $ f_{1}(a_{0},\cdots,a_{R_3}),\cdots, f_{8}(a_{0},\cdots,a_{R_3})$ be functions which   represent each term of $ P_{\rm SER}^{D}$.  For example, we assign    
\begin{equation}
f_{1} (a_{0})  = \frac{K_{1}}{a_{0}^{2(\mu_{S,D} + \mu_{S,R_1}+\mu_{S,R_2}+\mu_{S,R_3})}}, \, f_{2}(a_{0},a_{R_3}) = \frac{K_{2}}{a_{0}^{2(\mu_{S,D}+ \mu_{S,R_1}+\mu_{S,R_2} )}a_{R_3}^{2\mu_{R_3,D}}},\cdots
\end{equation} \label{AA} 
The second order derivative of $f_{1}(a_{0})$ w.r.t $a_{0}$ is given by
\begin{equation}
 \frac{\partial^{2}f_{1}(a_{0})}{\partial^{2} a_{0}} = \frac{4K_{1}(\mu_{S,D}+ \mu_{S,R_{1}}+ \mu_{S,R_{2}}+ \mu_{S,R_{3}} )(\mu_{S,D}+ \mu_{S,R_{1}}+ \mu_{S,R_{2}}+ \mu_{S,R_{3}}+\frac{1}{2})}{a_{0}^{2(\mu_{S,D}+ \mu_{S,R_{1}}+ \mu_{S,R_{2}}+ \mu_{S,R_{3}}+1)}}. 
 \end{equation}
 %
The Hessian matrix of $f_{2}(a_{0},a_{R_3})$, $\bigtriangledown^{2} f_{2}(a_{0},a_{R_3})$,  can be determined as  follows: 
\begin{equation}
H(a_{0},a_{R_3}) = \begin{bmatrix}\frac{4K_{2}(\mu_{S,D}+ \mu_{S,R_{1}}+ \mu_{S,R_{2}} )(\mu_{S,D}+ \mu_{S,R_{1}}+ \mu_{S,R_{2}}+\frac{1}{2})}{a_{0}^{2(\mu_{S,D}+ \mu_{S,R_{1}}+ \mu_{S,R_{2}}+1)}a_{R_3}^{2\mu_{R_3,D}}} & \frac{4K_{2}\mu_{R_3,D}(\mu_{S,D}+\mu_{S,R_1}+ \mu_{S,R_2})}{a_{0}^{2(\mu_{S,D}+\mu_{S,R_1}+ \mu_{S,R_2}+\frac{1}{2})}a_{R_3}^{2\mu_{R_3,D}+1}}\\ \frac{4K_{2}\mu_{R_3,D}(\mu_{S,D}+\mu_{S,R_1}+ \mu_{S,R_2})}{a_{0}^{2(\mu_{S,D}+\mu_{S,R_1}+ \mu_{S,R_2}+\frac{1}{2})}a_{R_3}^{2\mu_{R_3,D}+1}}& \frac{4K_{2}\mu_{R_3,D}(\mu_{R_3,D}+\frac{1}{2})}{a_{0}^{2(\mu_{S,D}+\mu_{S,R_1}+ \mu_{S,R_2})}a_{R_3}^{2(\mu_{R_3,D}+1)}} \end{bmatrix}.
\end{equation}
The principal minors of the  matrix $H(a_{0},a_{R_3})$, $H_{11}(a_{0},a_{R_3})\geq 0 $, $H_{22}(a_{0},a_{R_3})\geq 0 $ and $H_{11}H_{22} \geq H_{12}H_{21}$.  To this effect, the symmetric Hessian matrix $ H(a_{0},a_{R_3}) $ is positive semi-definite (PSD).  
  Since $\frac{\partial^{2}f_{1}(a_{0})}{\partial^{2} a_{0}} \geq 0 $ and $H(a_{0},a_{R_3}) $ is PSD, i.e.,  $\bigtriangledown^{2} f_{2}(a_{0},a_{R_3}) \succeq 0 $,  by the second order test in \cite{R9} both   $f_{1}(a_{0})$ and $f_{2}(a_{0},a_{R_3})$ functions are convex.  Following the same methodology,  it is shown that     the functions $f_{3},\cdots ,f_{8}$ are also convex.  Hence,  by the sum rule of convexity \cite[Section (3.2.1)]{R9} it follows  that the total function  $P_{\rm SER}^{D}$  is convex   w.r.t  $a_{0},a_{R_1},a_{R_2}\;\text{and}\;a_{R_3}$.

\vspace{-0.02cm}

\bibliographystyle{IEEEtran}

\end{document}